# Controlling hydrocarbon transport and electron beam induced deposition on single layer graphene: toward atomic scale synthesis in the scanning transmission electron microscope.

*Ondrej Dyck,[1] Andrew R. Lupini,[1] Philip D. Rack,[1,2] Jason Fowlkes,[1] Stephen Jesse[1]*

[1] Center for Nanophase Materials Science, Oak Ridge National Laboratory, Oak Ridge, TN

[2] Department of Materials Science and Engineering, University of Tennessee, Knoxville, TN

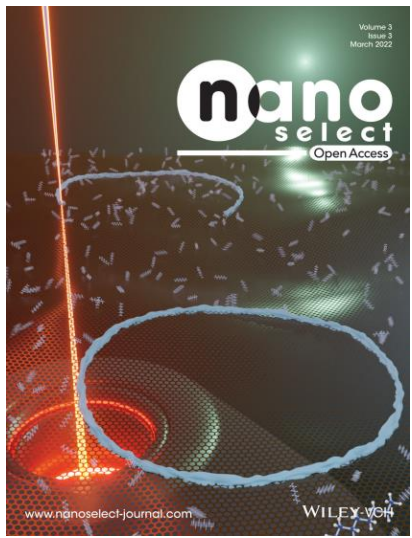

## Abstract

Focused electron beam induced deposition (FEBID) is a direct write technique for depositing materials on a support substrate akin to 3D printing with an electron beam (e-beam). Opportunities exist for merging this existing technique with aberration-corrected scanning transmission electron microscopy to achieve molecular- or atomic-level spatial precision. Several demonstrations have been performed using graphene as the support substrate. A common challenge that arises during this process is e-beam-induced hydrocarbon deposition, suggesting greater control over the sample environment is needed. Various strategies exist for cleaning graphene *in situ*. One of the most effective methods is to rapidly heat to high temperatures, e.g., 600 °C or higher. While this can produce large areas of what appears to be atomically clean graphene, mobile hydrocarbons can still be present on the surfaces. Here, we show that these hydrocarbons are primarily limited to surface migration and demonstrate an effective method for interrupting the flow using e-beam deposition to form corralled hydrocarbon regions. This strategy is effective for maintaining atomically clean graphene at high temperatures where hydrocarbon mobility can lead to substantial accumulation of unwanted e-beam deposition.

## Introduction

Focused electron beam induced deposition (FEBID) has been used to fabricate numerous 3D objects on the nanoscale.[1–8] Coupling FEBID capabilities with modeling of the physical processes involved provides a deeper understanding of the process and enables highly controllable 2D and 3D nanomanufacturing.[9–14] FEBID is typically performed in a scanning electron microscope (SEM) where an organometallic precursor gas (containing e.g., Pt, Au, or W) is injected across a substrate and dissociates with the focused electron-beam (e-beam). The dissociated molecular fragments bind to the substrate and to each other to form e-beam directed deposits that can be built up in a process akin to 3D printing on the nanoscale.

A question to ask is if such a technology can be ported to scanning transmission electron microscopes (STEMs) and exploit the increase in resolution due to the aberration-corrected probes often employed in STEMs. Whereas SEMs are used for visualizing the surfaces of bulk materials, aberration-corrected STEMs routinely resolve atomic columns in thinned materials and every atom in 2D materials.[15] Recent investigations have ventured into this territory, leveraging the finely focused STEM probe to induce atomic-scale adjustments to materials.[16–26]

Although aberration-corrected probes can be focused to sub-Ångstrom dimensions, interactions with the substrate are a significant contributing factor to the resolution of direct-write FEBID techniques due to the emission of secondary electrons (SEs).[27] Efforts to examine the role of substrate SE emission on FEBID in a STEM revealed substrate thickness-dependent growth rates, particularly in the early stages of nucleation where the majority of the material under the e-beam is the substrate.[28,29] By taking the idea of a reduced substrate thickness to the logical monolayer limit, one can envision working with graphene as the substrate[30] which is only one atom thick and is known to have a very low SE yield.[31–33]

Working in this regime, van Dorp et al. deposited nanodots from a $W(CO)_6$ precursor molecule-by-molecule to achieve deposits having dimensions less than 1 nm.[34] It becomes increasingly apparent, however, that extraneous material in the form of volatile hydrocarbons becomes a concern especially as the physical dimensions of the deposition site are edging toward the fundamental limit of single atoms. van Dorp et al. highlight the perniciousness of this challenge in their supplemental materials stating, "*For single-layer graphene exhaustive additional cleaning does not reduce the contamination level; contamination growth is visible when the graphene is observed at 800 °C in high vacuum (pressure at the sample better than $2\times10^{-7}$ mbar), in 2.8 mbar of nitrogen or in $3\times10^{-3}$ mbar of water, after plasma cleaning the microscope for 24 h and cleaning the holder for an additional 20 min.*"[34] Meyer et al. leveraged the presence of this contamination and used it as their precursor material to perform direct-write FEBID on graphene.[35] While this strategy cleverly works within the limitations imposed by less than pristine samples and may offer some select applications where carbon-based deposition is desired, it is not an ideal situation. The ubiquitous nature of hydrocarbon contaminants on graphene has not gone unnoticed and efforts to reduce it have been met with varied success.[36–40]

It is notable that hydrocarbon contamination in the STEM is observed in two configurations: 1) fixed to the sample surface and immediately imageable and 2) mobile hydrocarbons that form deposits under the e-beam that are not visible until after deposition. It is this second configuration, volatilized forms of hydrocarbons, that pose an impediment when trying to scale direct-write FEBID nanomanufacturing to the atomic scale. Diffusion along graphene or graphite surfaces is of fundamental interest in various other fields. The dynamics of aromatic hydrocarbons on graphene/graphite have been explored to gain a fundamental understanding of the processes involved.[41–46] These investigations have highlighted interesting properties such as ballistic diffusion[43,47] and the foundational principles governing friction at the atomic/molecular scale.[41,42,44,46] What is important for our discussion here is the model of gas adsorption onto a surface which presumes a gas phase, an adsorption layer, and the surface onto which the gas adsorbs.[45]

In this work we employ a rapid heat treatment to remove the fixed contamination from graphene surfaces and then explore the nature of the volatile form of hydrocarbons that deposit under the e-beam. We show that surface deposition can be used to arrest the ingress of additional hydrocarbons, suggesting that surface diffusion, rather than adsorption from vacuum, is the primary mode of hydrocarbon transport at high temperatures. We further present a reaction-diffusion modeling framework for describing the surface diffusion and deposition and discuss the interplay between precursor density and diffusion rate.

## Results and Discussion

### Sample overview and initial conditions

To investigate the behavior of hydrocarbons on a graphene surface, a suspended graphene sample was prepared on a Protochips Fusion™ heater chip using a wet transfer method (described in the methods section). Figure 1(a) shows a medium angle annular dark field (MAADF)-STEM image of the suspended graphene sample at room temperature (RT) and a depiction of the amorphous web of surface contaminants adhered to the graphene surface. The contaminants are stable in high vacuum (~$10^{-9}$ Torr) and presumably have reasonably long hydrocarbon chain lengths. Initially, no hydrocarbon deposition was observed at RT. The heater chip was then ramped to 900 °C at a rate of 1000 °C/s to remove the surface contaminants.[30,36]

Figure 1(b) shows a MAADF-STEM image of the sample after heating to 900 °C and a schematic drawing illustrating the observed changes. Large areas of graphene became clean as shown in the higher resolution MAADF-STEM image inset. Overview images of the sample before and after heating are shown in Figure S1. Residual contamination remains in some areas due to excessive e-beam exposure prior to heating as discussed elsewhere,[30,36] also shown in supplementary Figure S1. Upon decreasing the field of view to 64 nm (i.e., increasing the electron dose per unit area) rapid carbon deposition under the scanned area was observed. This is depicted in Figure 1(c), where a square area of carbon deposition appears in the center of the image as indicated by the outline/box. The schematic depicts deposition directed by the e-beam.

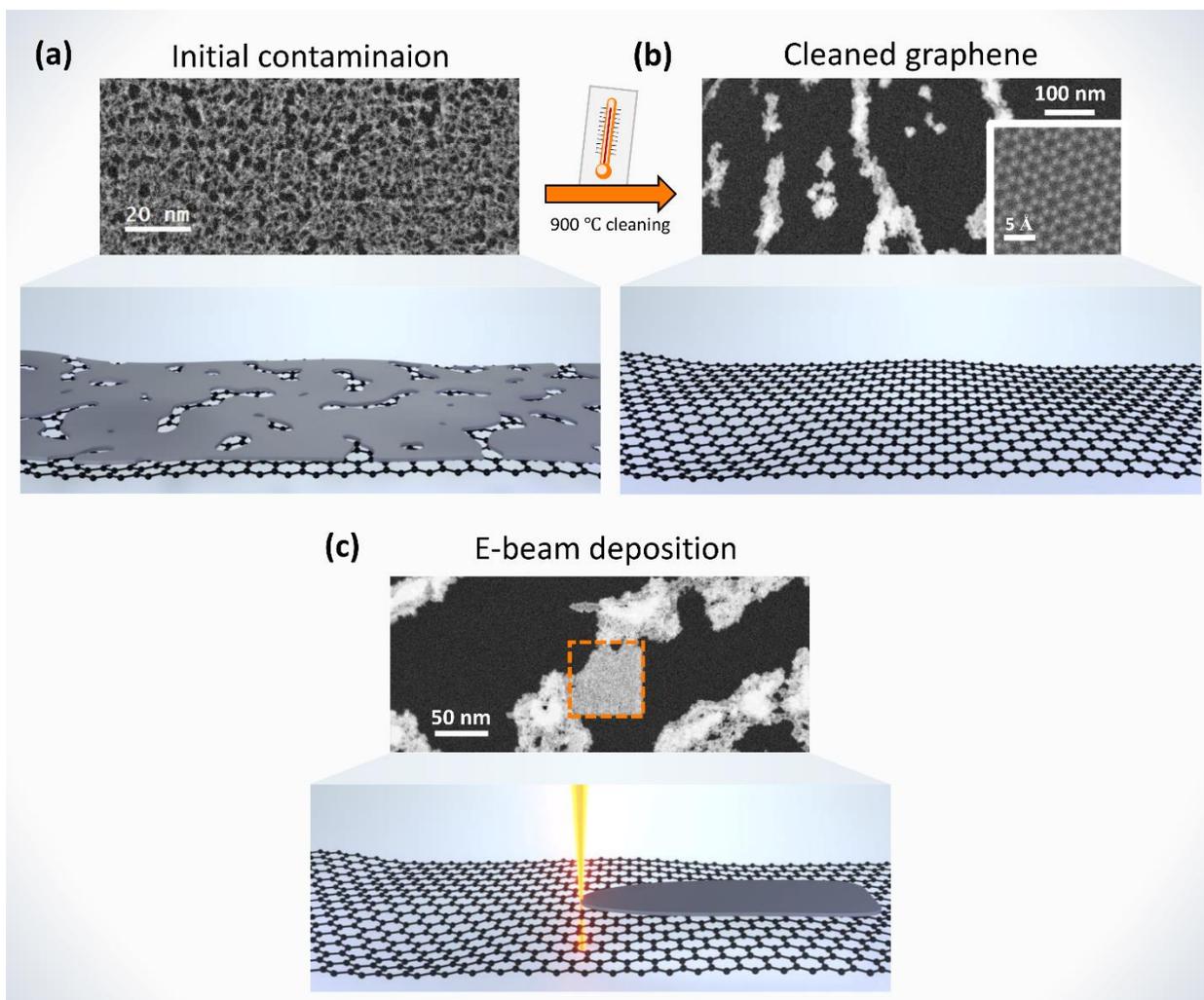

**Figure 1 Graphene sample overview.** (a) MAADF-STEM image of suspended graphene acquired at RT and accompanying schematic illustrating the amorphous contamination covering the graphene surface. (b) After ramping the heating chip to 900 °C at a rate of 1000 °C/s, large areas of the graphene are clean. Inset is an atomically resolved MAADF-STEM image of the clean graphene. (c) MAADF-STEM image of e-beam deposition performed within the box (outlined in orange). The area indicated was originally clean and zooming to a 64x64 nm field of view resulted in a square deposit where the e-beam was scanned.

This kind of deposition is quite pernicious as it grows under the e-beam and interferes with the imaging/processing of the region being investigated. This phenomenon, however, has been exploited as a direct-write nanoscale synthesis technique, as discussed in the introduction. Detailed Monte Carlo simulations of FEBID[9,48–50] shows that growth/etching proceeds via an interplay between the primary, secondary, and backscattered electrons with the precursor adsorbed in the e-beam interaction region. The supply of precursor to the e-beam interaction (raster) region is dictated by adsorption from the vapor phase as well as surface diffusion and is limited by the residence time of the adsorbed species. These transport mechanisms are all impacted by temperature, as recently studied in detail.[44]

While growth was not observed at RT, the e-beam clearly affects the region, dissociating or crosslinking the carbonaceous layer, since it remains intact after high temperature heating (center of Figure S1(b)). Large areas of the carbonaceous layer are removed during the fast heating in unexposed regions, as observed in Figure 1(b) and (c). Thermal dissociation of longer hydrocarbon chain, $C_xH_y$ molecules into smaller molecules is envisioned here. Subsequent carbon deposition at high temperature (Figure 1 (c)) suggests that transport of carbonaceous material occurs either by vapor phase adsorption or mobile and high-binding energy carbonaceous species that still reside on the surface.

*Examination of e-beam deposition*

To further illustrate high-temperature e-beam-induced deposition, a second example is shown in Figure 2(a-d). Figure 2(a) is an overview image of the locations where two 20-frame image stacks were acquired (indicated by orange dashed boxes). Figure 2(b-c) show the first and last frames from the stack (see SI for all image stack exposure e-beam scanning parameters). In **region i**, clean areas of graphene are not encircled by carbon walls and are termed an "open area." During scanning in **region i**, we observed hydrocarbon deposition along all the contamination edges as well as nucleation directly on the graphene surface in the open area (nucleation occurs at the three separate islands in Figure 2(b) just prior to initiating the capture of the data stack and a fourth nucleation site appears during acquisition). The areas of clean graphene from Figure 2(b) were artificially tinted orange with an opacity of 50% and overlaid on (c) to clearly highlight the locations where deposit growth occurred. Figure 2(d) shows a scatter plot of the linear growth rate as a function of electron dose obtained by analyzing the image stack, as described in the supplemental materials. On the right-hand side of the plot, a kernel density estimate (KDE) of the growth rate distribution is provided to illustrate the spread in measured growth rates within the observed data (see SI for additional details).

In contrast to the behavior observed in the **region i** "open area" shown in Figure 2(b-d), the second 20-frame image stack was acquired across an amorphous contaminant bridge that divides an "open area" forming an "enclosed area;" **region ii** is an area where the amorphous materials on the graphene surface completely surround a clean graphene area. Here, we note that growth only occurs on the open side and no growth occurs on the enclosed side of the bridge. The decrease in deposition rate was due to a global decrease in precursor availability over time as shown in supplementary Figure S3. The two datasets shown here were acquired 21 minutes apart.

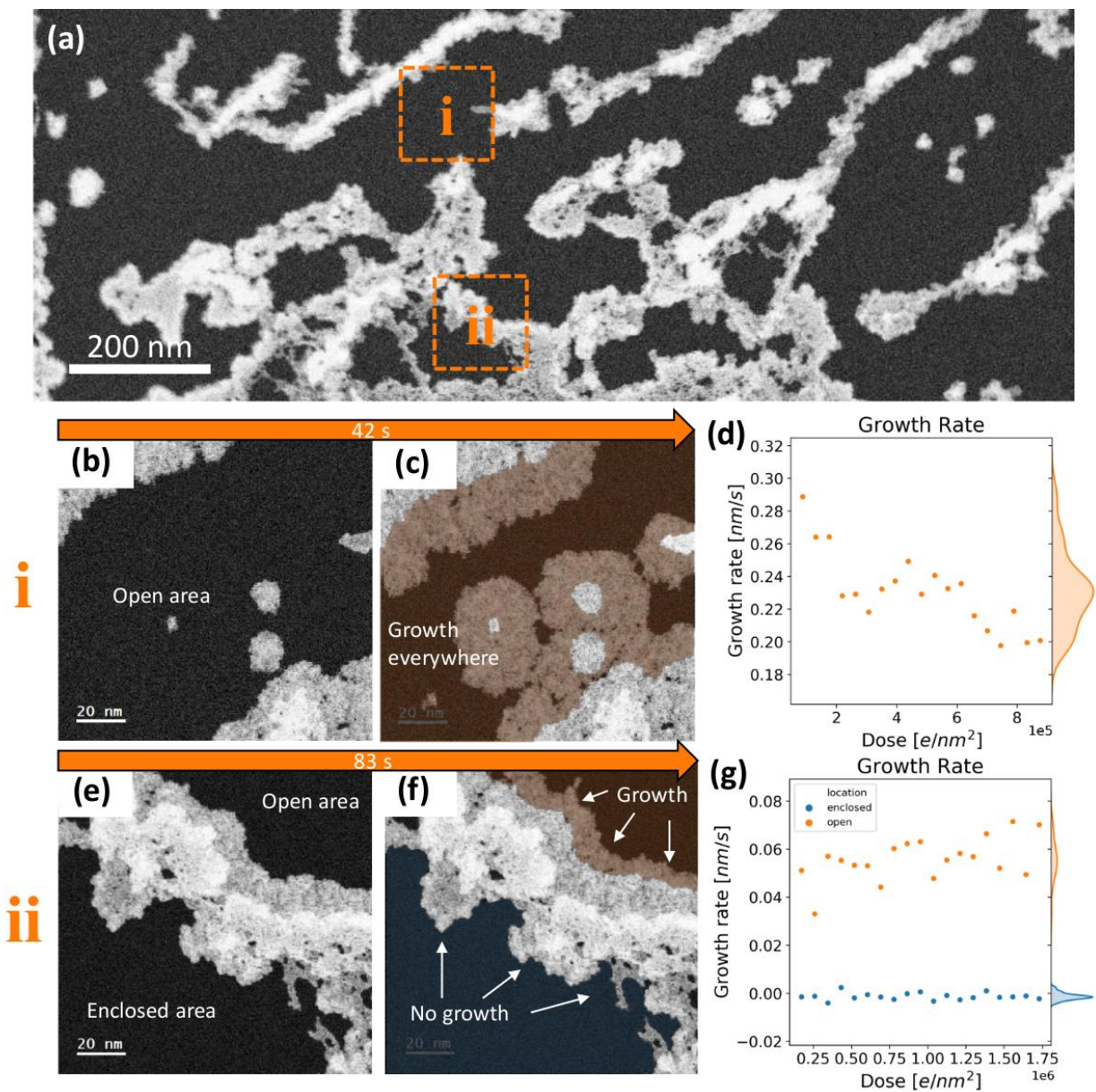

**Figure 2 Hydrocarbon growth in open areas vs. enclosed areas.** (a) Overview MAADF-STEM image showing the location where each image sequence was obtained. (b-c) **Region i** - first and last frame of a 20-frame image stack acquired over 42 s where growth occurred everywhere. As a guide to the eye, the clean area from (b) is tinted orange, set to 50% opacity, and overlaid on (c). (d) Extracted linear growth rate as a function of total electron dose. KDE of the growth rate distribution is plotted vertically on the right side. (e-f) **Region ii** - first and last frame from a 20-frame

image stack acquired over 83 s where an open region is adjacent to an enclosed region. Clean areas from e) are tinted orange for an open region and blue for an enclosed region, set to 50% opacity, and overlaid on (f) to highlight regions of growth. We observe no hydrocarbon growth in the enclosed region. (g) Extracted growth rate for each area.

No hydrocarbon growth was observed in the enclosed area, which suggests that migration of the hydrocarbons responsible for deposition is effectively blocked/trapped by the carbonaceous barriers observed on the surface. To confirm this, we used the e-beam to deposit a bridge of material on the graphene surface, cutting off a previously open region and making an enclosed region. Figure 3(a) shows the overview image prior to depositing the barrier. The orange dashed square, labeled region i, indicates the location chosen for deposition. The inset image shows the resultant barrier after deposition. After depositing the barrier, a 20-frame image stack was acquired to document where hydrocarbon deposition occurred after construction of the barrier, as shown in Figure 3(b-d). Deposition occurred only in the open area, as observed in Figure 3(c). Figure 3(e-g) shows a repeat of the experiment from Figure 2(e-g); however, since both sides of the amorphous bridge are enclosed areas, we observe no deposition on either side of the bridge.

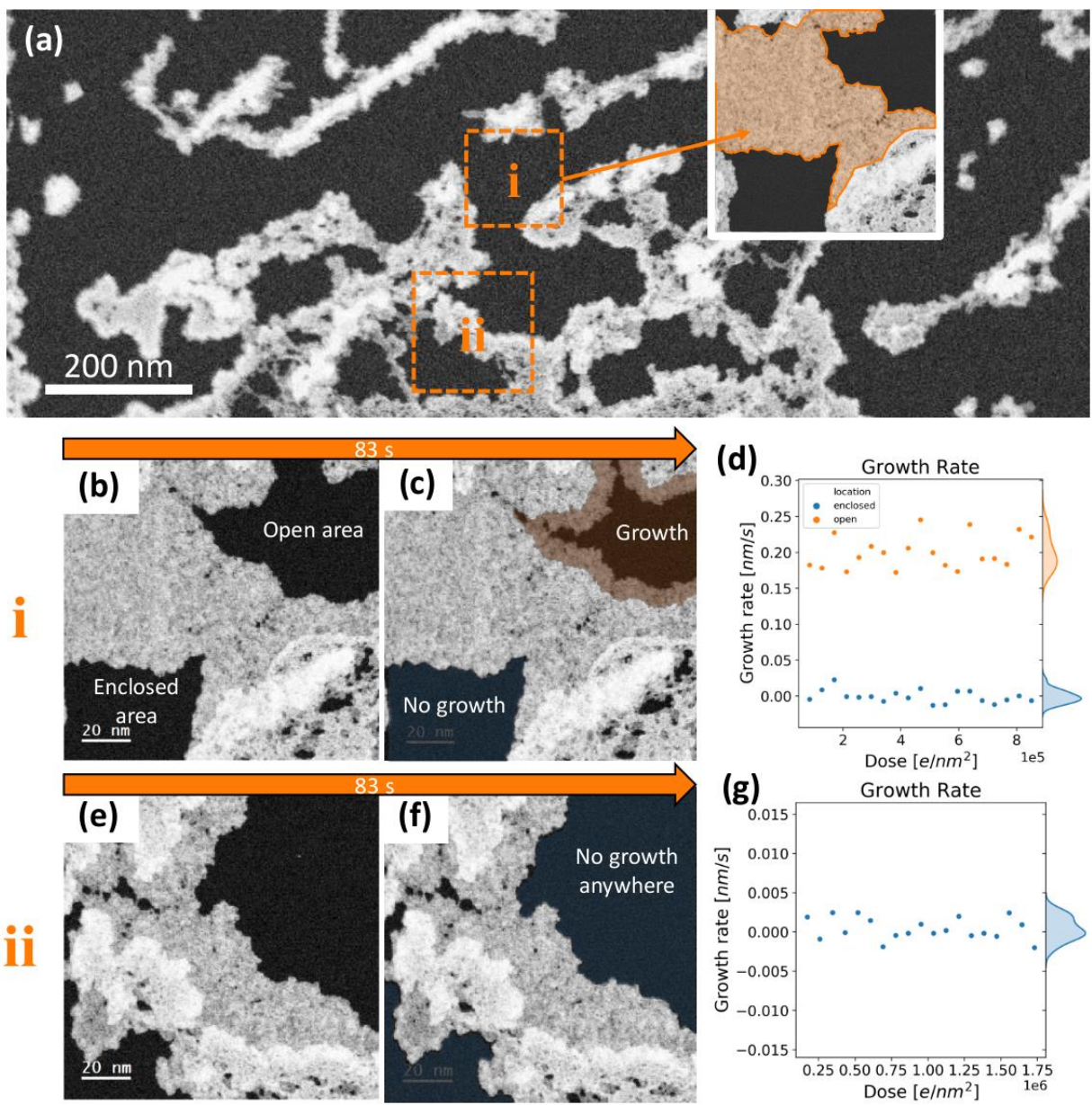

**Figure 3 Demonstration that an enclosure prevents hydrocarbon growth.** (a) Overview MAADF-STEM image showing the location of the deposited barrier and locations for subsequent acquisition of image stacks. Inset is an image of the deposited barrier. (b-c) after the barrier was constructed a 20-frame image stack was acquired of the barrier showing the difference between the open area (region i) and enclosed areas. Clean areas of graphene from (b) are tinted orange and blue, set to 50% opacity, and overlaid on (c) to highlight where growth occurred. (d) Extracted linear growth rates in each location. (e-f) Repeat of the experiment shown in **Figure 2**(e-g) where all the graphene within the field of view is enclosed and no deposition is observed. Clean areas from (e) are tinted blue, set to 50% opacity, and overlaid on (f) to illustrate that no change occurred. (g) Extracted linear growth rate.

To test whether this strategy can work on a larger scale, we identified a region that was almost enclosed but exhibited heavy deposition. We used the deposition to form a barrier of material and generate a relatively large, enclosed region. Figure 4(a) shows an overview image prior to construction of the barrier with the targeted enclosed area highlighted in blue and with the desired position of the barrier shown by the dashed orange line. The image inset shows the barrier after it was deposited. A 20-frame image stack was acquired to observe the growth on either side of the barrier, as shown Figure 4(b-d). The enclosed side of the barrier shows no deposition despite the size of the enclosed area and the high deposition rate prior to forming the enclosure (e.g., the barrier itself was constructed using the deposition that occurred in this area).

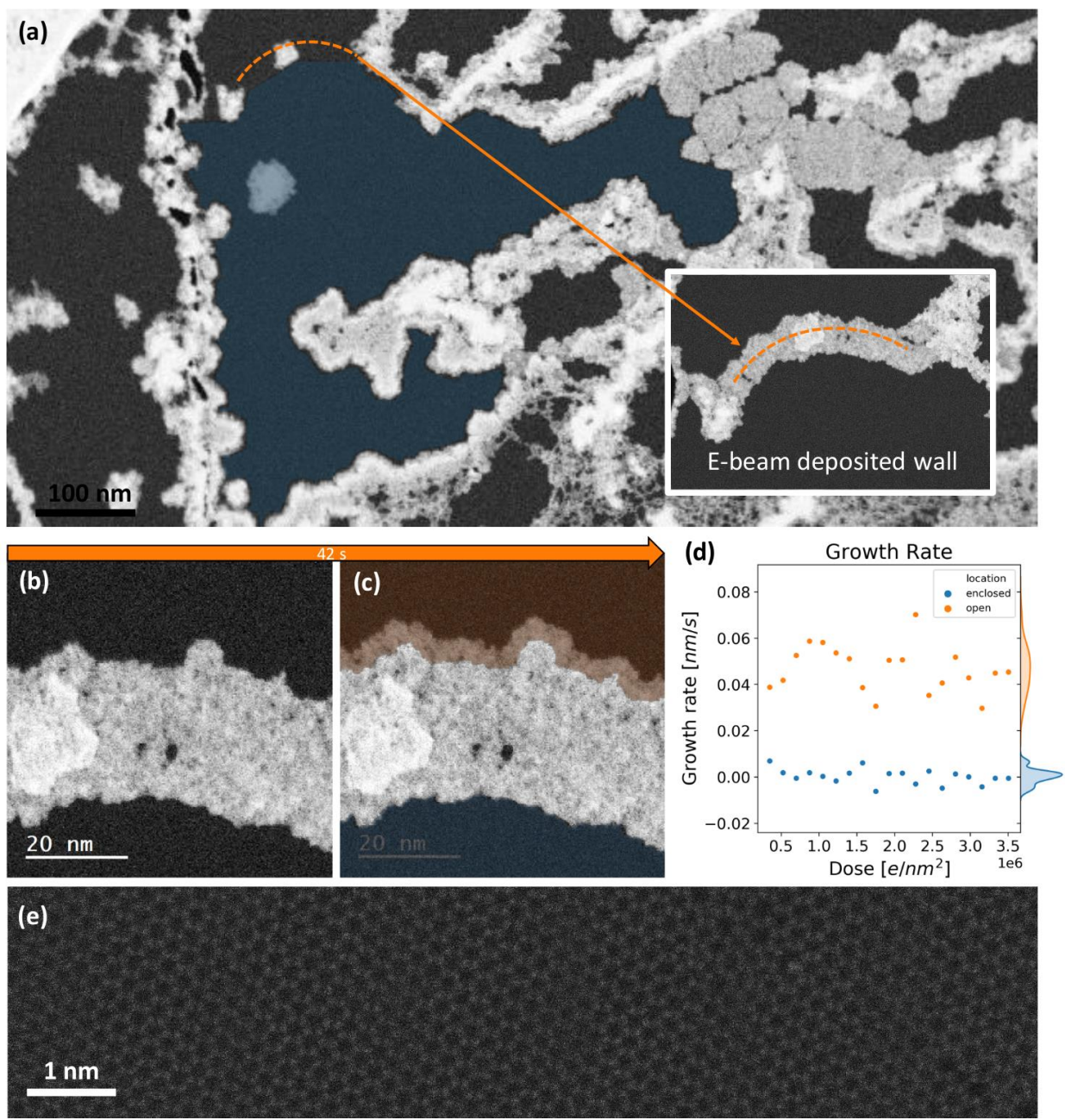

**Figure 4 Creating a large, enclosed area.** (a) Overview MAADF-STEM image of the area to enclose (highlighted by the blue overlay). The chosen location for depositing a hydrocarbon barrier is indicated by the orange dashed curve. Inset is an image acquired after deposition of the barrier. (b-c) First and last frames from a 20-frame image stack acquired across the barrier. Clean areas in (b) are tinted blue and orange, set to 50% opacity, and overlaid on (c) to highlight growth locations. Growth only occurs outside the barrier. (d) Extracted linear growth rate of both locations. (e) MAADF-STEM image of an area of atomically clean graphene. Prior to construction of the barrier, acquisition of such an image was not possible.

## *Discussion*

Regarding the transport mechanism responsible for the growth, adsorption from the vapor phase can be ruled out as the enclosed regions would also experience growth. The $10^{-9}$ Torr base pressure is also prohibitively low to be the dominant precursor supply. Thus, surface diffusion must be the primary mechanism, which is evidently "blocked" by the deposited carbon. Interestingly, this suggests that the diffusing species has a high adsorption energy, otherwise it would quickly desorb at 900 °C. Regarding the blocking mechanism, questions regarding the diffusing species are 1) do they get trapped at/in the carbon barriers, 2) do they arrive at the carbon barriers and desorb, or 3) do they arrive and effectively reflect from the carbon barrier? If the diffusing species arrive and become trapped in the carbon deposits, then one would expect a coarsening of the carbon contamination regions everywhere at high temperature, which would be independent of e-beam exposure. If the diffusing species arrive and desorb, then one would expect the carbon growth rate to decay rapidly as the high mobility of the species would eventually exhaust the surface concentration. The deposition rate decays with time but deposition did continue for over an hour (see Figure S3), which suggests that if some desorption occurs at the edges it is not the primary interaction mode. Thus, the carbon barriers primarily reflect the mobile hydrocarbon species on the surface. Finally, regarding growth in the open regions, two scenarios are possible as high temperature growth appears to occur preferentially at the barriers, primarily through periodic nucleation, likely at vacancy sites created by knock-on damage (see Figure 2(b-c)). First, the e-beam could activate the diffusing species that can subsequently diffuse and stick to the sidewall of the barrier. Second, the e-beam-induced growth could be enhanced at the barrier sidewall due to enhanced e-beam interactions with the thicker carbon layer leading to more SE emission. It is likely that there is a combination of these two effects.

## Modeling

A numerical simulation was developed to ascertain whether a reaction–diffusion mechanism could be responsible for determining the lateral film deposition rates observed for high-temperature STEM e-beam–induced deposition experiments conducted on graphene. The fact that the chemical composition of the carbon–based precursor is unknown presents a significant challenge to constructing a *detailed* model since many critical parameters that can influence deposition are unknown: the unknown parameters include the precursor surface diffusion coefficient, the equilibrium precursor surface concentration, and the electron impact dissociation cross-section of the precursor molecule. For this reason, a simple model was constructed to determine if the proposed mechanism is at least plausible, i.e., by using physically realistic ranges for unknown parameters, could the experimental deposition rate be accounted for?

*Overview*

To test the validity of the reaction–diffusion model, a set of physical parameters was sought that would predict the lateral film growth rate for a set of real experiments. Specifically, two parameters were varied over a multi–decade range of values to find a solution, namely the precursor surface diffusion coefficient (D) and the equilibrium precursor surface concentration ($C_o$). Note that ($\theta$) will be used in place of ($C_o$) when reporting simulation results where the parameter ($\theta$) represents the equilibrium precursor surface coverage for the beam–off condition; it is the fraction of the monolayer coverage of 60 sites/nm$^2$, which is roughly the surface density of graphene atoms.

(D) and ($C_o$) were selected to vary because they appear to have significant roles in determining experimental outcomes as rationalized by comparing experiments; reasonable guesses were made for additional unknowns as described below. Simulations revealed a range of (D and $C_o$) combinations for each experiment that reproduced the observed experimental growth rates. However, the proposed mechanism may only be deemed as a potential candidate – a unique set of parameters was *not* identified that described all of the experiments. Nonetheless, if one considers the possibility of a decreasing vapor pressure during the experiments then a range of solution sets tend to align for multiple experiments.

*Reaction–Diffusion Model for STEM EBID*

A 2D numerical simulation of time–dependent precursor surface diffusion and e-beam-induced dissociation/deposition (EBID) was constructed using the fully implicit finite difference method to emulate actual STEM experiments. The Dirichlet boundary condition is applied at three of the four scanning frame boundaries (--) while a mass conservation balance is applied at the film/substrate interface. The former boundary condition was defined by the observation that EBID conducted in open regions yields a nearly continuous linear growth rate during STEM experiments suggesting a stable reservoir of precursor exists on the graphene surface while the latter boundary condition reflects the experimental observation that deposition is initiated at a preexisting interface and seems to depend strongly on the precursor surface diffusion. Thus, e-beam-induced dissociation/deposition only takes place at regions of contact with the film interface. Importantly, interface transport is emulated in the simulation by displacing the origin of the e-beam (in response to deposition), which makes it possible to keep the simulation domain effectively stationary, as summarized in Figure 5(a–b). The mathematical aspects of the simulation are described briefly while a more comprehensive description is provided in the supplemental information.

The 2D surface diffusion equation is solved for a rectangular domain represented schematically in Figure 5(a) as colored (●) elements

$$\frac{\partial C}{\partial t} = D\nabla^2 C(x,y) \quad [i]$$

A mass conservation balance is applied at the film/substrate interface (the left-most vertical column of nodes) as

$$\frac{\partial C}{\partial t}\frac{\Delta x_o}{2}\Delta y = D\frac{\partial C}{\partial x}\bigg|_{+\delta x}\Delta y + D\frac{\partial C}{\partial y}\bigg|_{+\delta y}\frac{\Delta x_o}{2} + D\frac{\partial C}{\partial y}\bigg|_{-\delta y}\frac{\Delta x_o}{2} - \iota\sigma i''_{SE} C \frac{\Delta x_o}{2}\Delta y \quad [ii]$$

where the first three terms on the right–hand side describe precursor surface diffusion and the last term accounts for the dissociation of precursor at the interface. The parameters ($\iota$) and ($\sigma$) are constants for all experiments. ($\iota$) represents the number of deposited film molecules produced per precursor molecule dissociated. ($\iota$) was set to 1 because the precursor molecule chemistry is unknown while ($\sigma$) was set nominally to $10^{-2}$ nm$^2$, a common order of magnitude for EBID. The SE surface density, or ($i_{SE}$''), was simply set equal to the primary e-beam current density without knowledge of the SE yield on the film. Note that $i_{SE}$'' is a function of (x), (y), and (t) as the beam is scanned through the frame but deposition is possible only at the current film interface position.

The lateral film deposition velocity ($v_f$) is described by

$$v_f = \iota\sigma i''_{SE}\frac{\Delta x_o}{2z_o}\frac{C}{C_f} \quad [iii]$$

where ($C_f$) is the deposited film concentration in units of molecules *per unit volume of film*, ($\Delta x_o$) is the simulation pixel width at the interface boundary, ($z_o$) is the film thickness (taken here as the characteristic molecular size 0.3 nm), and (C), as a reminder, is in units of molecules *per unit area of surface*.

*Reaction–Diffusion Simulation for STEM EBID*

The STEM square scan frame, shown from the point–of–view of the incident e-beam (see gray box), is shown in Figure 5(a) as a reference frame to explain the equivalent simulation frame–of–reference as well as to indicate key simulation aspects.

The primary e-beam executes a raster motion beginning with a linear scan in the x–direction, from left–to–right. Upon completion of a single line scan, the beam is displaced by one pixel in the positive y–direction and the process repeats. A frame is completed when the entire square is scanned. The frame edge dimension is defined as ($y_{ROI}$).

The simulation emulates real frame scanning using a modified reference frame, i.e., a frame that is fixed to the interface position per pixel row. This choice of reference frame ultimately leads to a moving and distorted simulation frame over time (see hatched, black frame). This frame of reference makes it possible to fix the simulation pixels in space: stationary simulation spatial nodes (see colored nodes) simplify simulation scripting and minimize the number of calculations required.

Also note, as a further means to improve time efficiency, the node spacing increases in the scanning direction – at relatively large distances from the interface, surface gradients in C(x) are smaller, allowing for a larger step size and thus fewer computational nodes. The initial ($\Delta x_o$) and final ($\Delta x_f$) step sizes are reported in the SI 2. The step size in the y–dimension is a constant parameter ($\Delta y$).

Lastly, computation time is also minimized by advancing the primary e-beam to the starting position of the next scanning row when the precursor surface coverage in the entire frame domain returns to the equilibrium value ($C_o$). The primary e-beam is modeled as a Gaussian probe (FWHM = 2 nm) so; in the limit of a small electron probe size relative to the frame size, negligible deposition occurs when the beam moves several nanometers beyond the interface allowing for the replenishment in the frame by precursor surface diffusion from the frame boundaries. This fact explains why the simulation frame width ($x_{ROI(v)}$) is smaller than the actual experimental frame size in the schematic in Figure 5(a).

The moving reference frame requires the definition of two beam shifts per row; (1) an initial shift $x_{b,o}(y)$ that equals the distance between the left edge of the frame and the initial contamination/graphene interface position and (2) the evolution of lateral film growth in the x–dimension with time $x_f(y,t)$. Figure 5(b) shows this shift pair for a particular row in a real STEM image as a visual example of the mathematical shift. The left–most STEM image shows the experimental reference frame while the rightmost image shows the simulation frame. The shift variables for a single row are shown superimposed over the simulation frame for clarity in Figure 5(a).

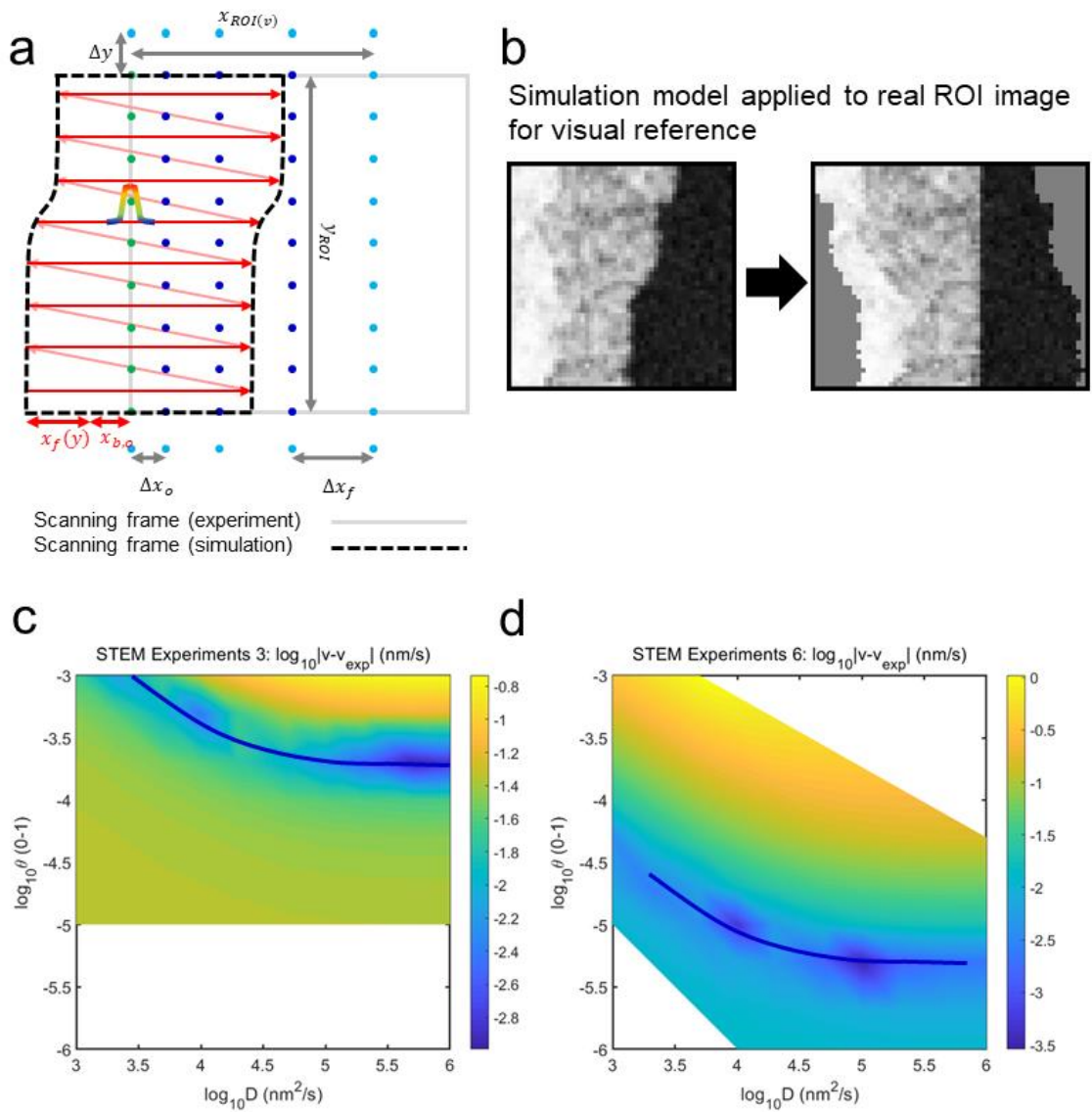

**Figure 5** (a) Schematic of the relationship between the computation pixel array (●), the real e-beam scanning frame (-) and the simulated e-beam scanning frame (--). (b) Example STEM image in a frame scan during deposition (left–most image). The brightest region represents the original surface contamination that existed prior to EBID, the intermediate brightness shows deposited film, and the dark region is exposed graphene. Right–most image shows the same image but shifted into the simulation frame of reference. (c) and (d) show results for two cases where the surface plots represent the difference in the lateral film growth rate between experiments and simulations. Estimates for the solution curve are superimposed.

## *Simulation Results*

Simulations representing four real experiments were conducted. In each case, the parameters ($\theta$) and (D) were both varied over nominally three–orders of magnitude. Multiple solutions were found for each experiment, which traced out a curve in solution space. These solutions predict the experimental lateral deposit rates. Figures 5(c–d) show results for two cases where the surface plots represent the difference in the lateral film growth rate between experiments and simulations. The color represents the degree of agreement between the experimentally observed growth rate and the model prediction across a range of values for ($\theta$) and (D). All axes, including the colormap, are base 10 logarithmic. Note the blue curves are superimposed estimates of the solution curve. The color variation in the surface map along the solution curve results from data interpolation. The non–uniformities are caused by variations in data sampling density as well as the addition of gradient descent execution. The surface maps for all four experiments are provided in the SI.

## *Discussion*

Several characteristics of the simulated data are relevant to the current problem. First, the approximate shape of the solution curve is common for all four experiments. For example, in Figure 5(c), although the curve was qualitatively traced to suggest the path of the true solution curve, translation of the curve to Figure 5(d) reveals the same solution curve shape. On the one hand the simulated results suggest that the deposition model is a potential solution for all experiments; however, on the other hand the best solution seems to shift in parameter space for each experiment, i.e., Figure 5(c) versus Figure 5(d), suggesting the model may not predict a unique solution that is valid across all experiments.

The physical deposition regimes are described with reference to the solution space. Specifically, the shape of the (D,$\theta$) solution curve for any given experiment is related to the deposition regime, i.e., whether deposition is diffusion-controlled or dissociation-controlled. Consider Figure 5(c) where at relatively lower values of (D), (θ) varies with (D). Surface diffusion controls deposition in this region of the (D,θ) parameter space, e.g., a decrease in (D) requires a concomitant increase in (θ) to sustain the deposition rate. Conversely, at larger (D) values, changes in (D) require negligible changes in (θ) because the reaction is electron limited. Here, changes in the e-beam current density would require changes in (θ) to sustain a constant growth rate.

Interestingly, the solution curve can be shifted solely in (θ) to overlay the curve with the solution map for all four experiments. Further, it was observed that the lateral film deposition rate decreased as a function of time after sample heating in the vacuum chamber, or ($\Delta t_{vac}$) (SI 3). Without the ability to control the total precursor pressure during the experiments, it is expected that the surface concentration of precursor should decrease in time. Thus, a steady decline in the total precursor concentration could explain the (-Δθ) shifts of the solution curves. In further support of this solution is the fact that (-Δθ) increases as ($\Delta t_{vac}$) increases.

In summary, simulations and experiments suggest the following characteristics of high-temperature, in situ STEM deposition;

1. the precursor molecules must be relatively large to remain attached to the surface at the elevated deposition temperature;
2. the precursor molecules must have a relatively low concentration in the vapor phase since closed regions exhibit no detectable deposition;
3. the characteristic precursor desorption time must be on the order of at least hours, which is also consistent with a relatively large precursor molecule.

Postulates 1–3 combined suggest that the deposition rate depends primarily on the surface diffusion of the precursor.

4. Surface diffusion rates must be large in magnitude to sustain growth in open regions if the local surface concentration is as low, as suggested by postulate 2. (In fact, a relatively large surface diffusion rate is expected at the elevated temperatures used in the experiments.)

In favor of the postulated model is the self–consistency of statements 1–4, which are void of contradiction.

## Conclusion

These results deepen our understanding of the behavior of hydrocarbons on the surface of graphene and show a route to preserve atomically clean areas under conditions where substantial hydrocarbon deposition was previously observed. The strategy employs the use of a physical barrier attached to the graphene surface, which is effective at blocking the flow of C adatoms and hydrocarbons. The effectiveness of this technique indicates that C adatom and hydrocarbon diffusion occur almost entirely along the graphene surface with a limited role for gaseous hydrocarbons in the vacuum being dissociated by the e-beam and attaching to the graphene. We have demonstrated *in situ* e-beam deposition of physical barriers to mitigate the influx of hydrocarbons; however, *ex situ* strategies may also be amenable to similar processing. For example, e-beam lithography or electron/ion beam deposition may be employed to pattern long range barriers to block ingress of unwanted contaminants. Combined with previous advances in *in situ* STEM nanofabrication, this strategy for maintaining atomic cleanliness of graphene during e-beam processing holds promise for facilitating atomic scale e-beam nanomanufacturing.

## Methods

### *Sample preparation*

Graphene was grown on Cu foil using atmospheric pressure chemical vapor deposition (APCVD)[40] and spin coated with poly(methyl-methacrylate) (PMMA) to protect the graphene and act as a mechanically stabilizing membrane. The Cu/graphene/PMMA stack was floated on a solution of ammonium persulfate and deionized (DI) water (0.05 g/ml) until the Cu foil was etched away. The sample was then transferred to a DI water bath to rinse away residual ammonium persulfate. The sample was caught on a Protochips™ Fusion heater chip and baked on a hot plate at 150 °C for 15 minutes to promote adhesion of the graphene to the chip. After cooling, the PMMA was dissolved using acetone and the chip was dipped in isopropyl alcohol to remove the acetone residue and dried in air. To further clean the sample[36] the chip was put in vacuum and ramped using the built-in heater to 1200 °C at a ramp rate of 1000 °C/ms. The sample was then stored in air. Just prior to examination in the STEM, the sample was loaded into the STEM holder cartridge and baked at 160 °C in vacuum for 10 hours to remove adsorbed moisture and light hydrocarbons.

### *STEM imaging*

A Nion UltraSTEM 200 was used for these experiments. It was operated at an accelerating voltage of 100 kV, a nominal beam current of 60 pA (except for the image acquired in Figure 4(e) where a decreased beam current of 20 pA was used to increase resolution), and nominal convergence angle of 30 mrad. All images were acquired using a medium angle annular dark field (MAADF) detector.

## Acknowledgements

This material is based upon work supported by the U.S. Department of Energy, Office of Science, Basic Energy Sciences, Materials Sciences and Engineering Division and was performed at the Oak Ridge National Laboratory's Center for Nanophase Materials Sciences (CNMS), a U.S. Department of Energy, Office of Science User Facility.

# Controlling hydrocarbon transport and electron beam induced deposition on single layer graphene: toward atomic scale synthesis in the scanning transmission electron microscope.

*Ondrej Dyck,[1] Andrew R. Lupini,[1] Philip Rack,[2] Jason Fowlkes,[1] Stephen Jesse[1]*

[1] Center for Nanophase Materials Science, Oak Ridge National Laboratory, Oak Ridge, TN

[2] Department of Materials Science and Engineering, University of Tennessee, Knoxville, TN

## Supporting Information 1

*Graphene Sample Overview*

**Figure S1** shows an overview of the sample used in these experiments. As described in the methods section of the main text, graphene was transferred from its Cu foil growth substrate onto a Protochips™ Fusion heater chip and examined in a Nion UltraSTEM 200 using an accelerating voltage of 100 kV. **Figure S1**(a) shows an overview medium angle annular dark field (MAADF) STEM image of the sample at room temperature prior to any heating. The graphene is observed suspended over a circular aperture in the heater chip substrate (the substrate appears bright around the edges of the image). Substantial surface contamination was observed on the suspended graphene and a magnified image of this contamination is shown as an inset. The overlaid text and arrows indicate a few example regions (the uniform darker gray areas) which correspond to the contamination morphology shown in the magnified view. The sample temperature was then ramped to 900 °C at a rate of 1000 °C/s and held at that temperature. **Figure S1**(b) shows an overview image after this process (at 900 °C). The surface contaminants indicated in (a) have desorbed, particularly along the edges of the aperture. Initial imaging of the sample was performed at the center of the suspended region and exposure of the contaminants to the electron beam (e-beam) has caused them to adhere more strongly to the graphene surface. This is the reason that less "cleaning" action is observed toward the center of the image. The darker areas indicated by the text and arrow overlay correspond to atomically clean areas of graphene. The inset shows an image obtained in one of these areas where we can clearly see the hexagonal single layer graphene lattice. The image in **Figure S1**(c) was acquired in the region indicated by the box in (b). At the center of this image appears a square of carbon deposited by the e-beam scanning over the boxed region as indicated by the overlaid text and arrow. This illustrates that while these areas appear to be atomically clean, mobile hydrocarbons diffuse under the e-beam and adhere to the surface. Thus, there exists some concentration of surface molecules that are not visible in the images.

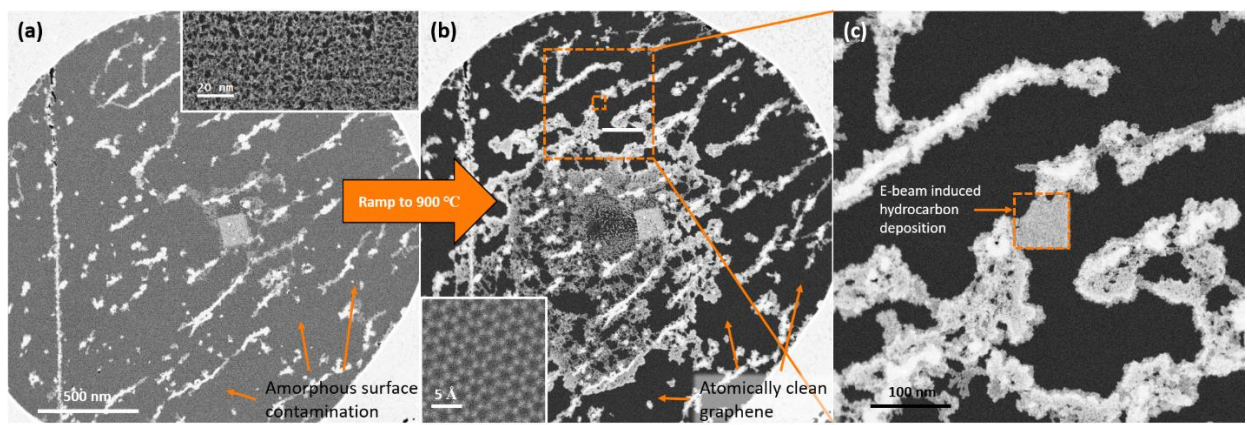

**Figure S1 Graphene sample overview**. (a) MAADF-STEM image of the suspended graphene acquired at room temperature and accompanying schematic drawing illustrating the amorphous contamination seen covering the graphene surface. Inset is a magnified view. (b) After ramping the heating holder to 900 °C at a rate of 1000 °C/s, large areas of the graphene become atomically clean. Inset is an atomically resolved image of the clean graphene. (c) higher resolution image of the area within the large box in (b). The area indicated at the center was originally clean and zooming in there resulted in a square of deposition where the beam was scanning.

## *Deposition Rate Analysis*

In Figures 2-4 of the main text we show growth rates extracted from the accompanying STEM videos. The procedure used for extracting these growth rates is described here.

To calculate growth rates from sequential STEM images (videos) the trainable Weka segmentation plugin[1] for ImageJ/Fiji[2] was first used to create a segmentation mask to distinguish between areas of clean graphene and contamination. **Figure S2**(a) shows a typical image frame and **Figure S2**(b) shows the same frame with the segmentation mask color coded and overlaid. The number of pixels belonging to the contaminated regions were then counted and converted to area. This procedure allowed a reliable measure of areal growth (i.e. square nanometers covered per unit time or dose), shown in **Figure S2**(c). To extract a linear growth rate perpendicular to the growth front, the following procedure was used: edges were first detected in the segmentation mask video using Canny edge detection,[3] contours were then found using an algorithm developed for topological structural analysis.[4] These tools were used as implemented in the python OpenCV package.[5] Edge contours were rank ordered according to the level of nesting detected (e.g. contours contained within contours) and their pixel length logged. Contours which were not at the top level or had lengths less than 40 pixels were rejected to prevent inflation of the measured edge length from the addition of short, spurious edges (i.e. noise). **Figure S2**(c) shows the areal coverage as a function of time and dose. The increase in areal coverage, $\Delta A_i$, between consecutive frames, $F_i$ and $F_{i+1}$, was then divided by the total edge length, $L_i$, detected in frame $F_i$, providing a measure of linear growth rate through time, shown in **Figure S2**(d). A kernel density estimate was calculated based on the observed linear growth to provide a visual representation of the estimated distribution of values. This is plotted vertically to the right of the scatterplot. A gaussian kernel was used for smoothing and Scott's Rule was used for bandwidth selection.[6]

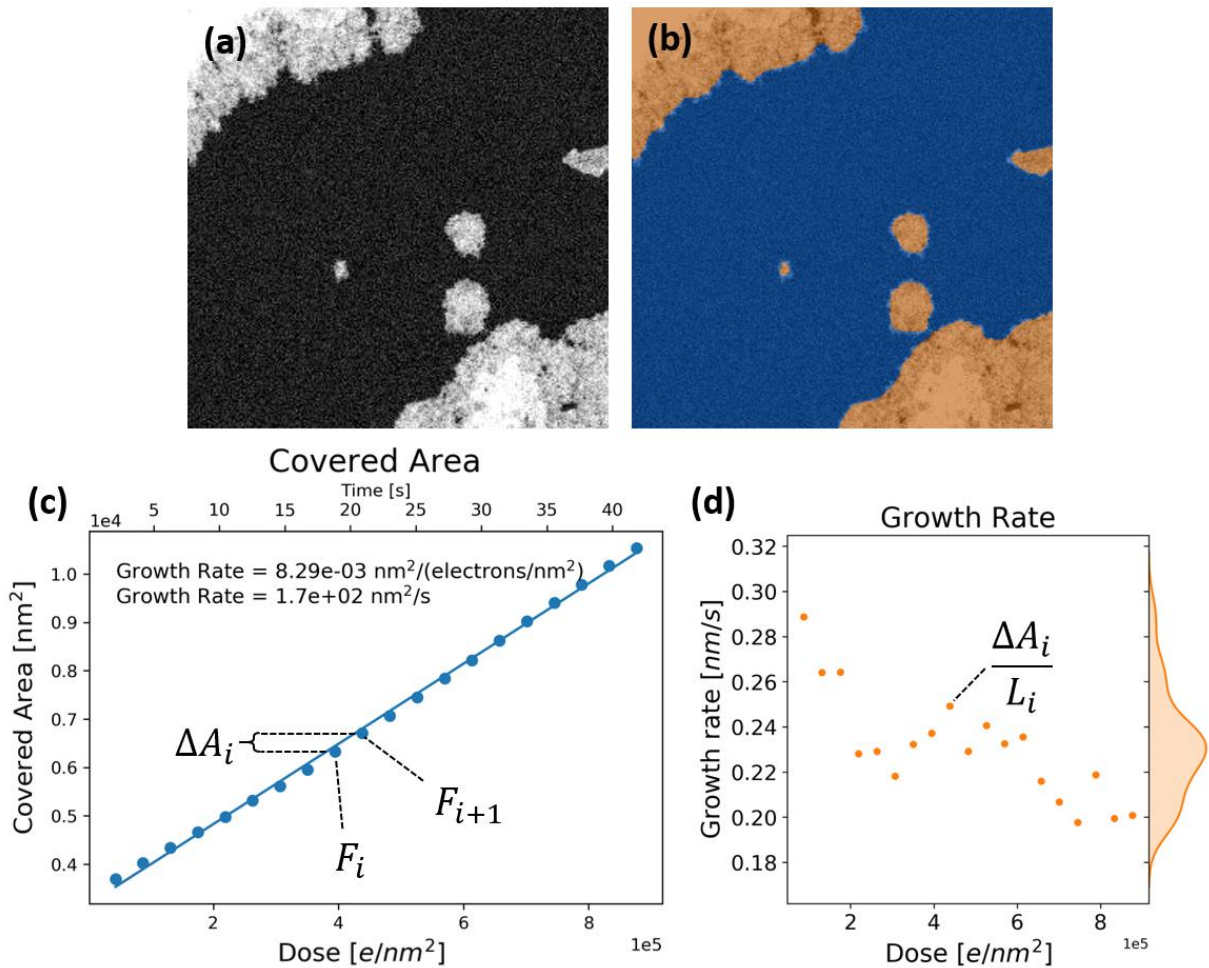

**Figure S2 Summary of growth analysis.** (a) An example frame from one of the STEM videos. (b) The same frame from (a) with the segmentation map color coded and overlaid. Edge contours were extracted from the binary segmentation map using Canny edge detection implemented in OpenCV. (c) A plot of the covered area as a function of time (upper x axis) and dose (lower x axis). Circles represent experimental measurements and the line is a linear fit to the measurements. The slope of the linear fit is overlaid giving a measure of areal growth rate. Linear growth rate was calculated by taking the change in covered area, $\Delta A_i$, between two sequential frames, $F_i$ and $F_{i+1}$, and dividing by the edge length measured in the i$^{th}$ frame, $L_i$. These values are plotted in (d) as a function of dose. A kernel density estimate of the growth rate distribution is plotted on the vertical axis to the right to provide a qualitative visual for the spread of the underlying distribution.

To capture the deposition rate as a function of time several STEM data sets were acquired spanning over an hour. The average growth rate observed within each data set is shown in Figure S3. An exponential fit was performed and is shown as a line. Fit parameters and equation are overlaid.

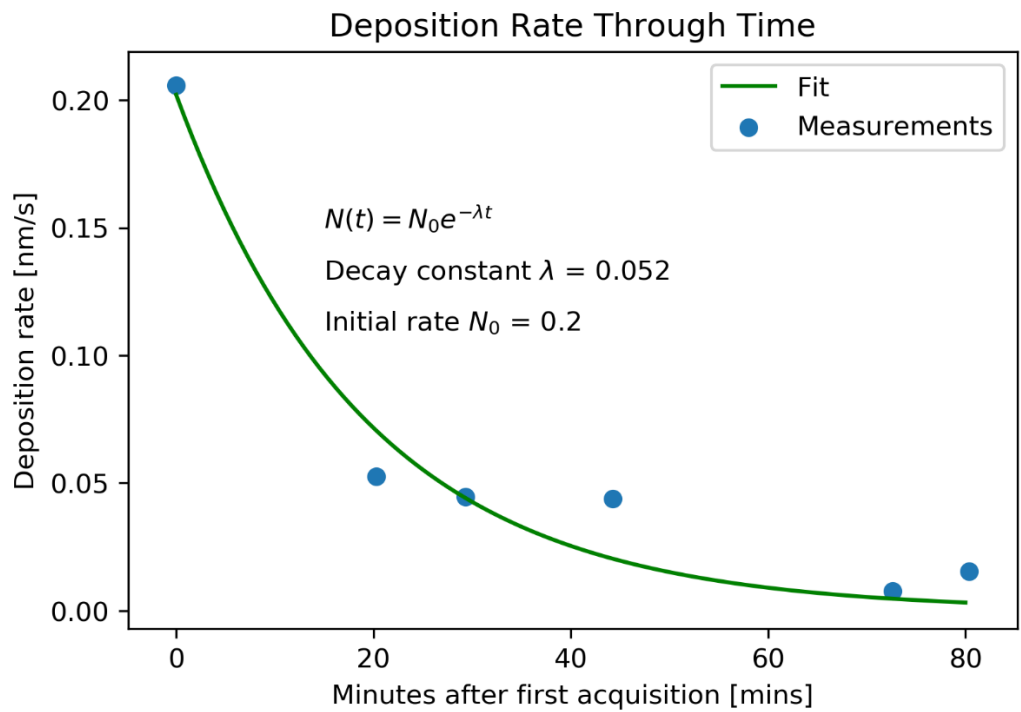

**Figure S3: Average deposition rate observed over time.** Dots represent the average rate observed in a given STEM data set. Line represents an exponential decay fit to the data points shown. Equation and parameters of fit are overlaid.

## *Experimental Parameters*

Table S1 summarizes the parameters used to acquire each STEM video presented in this manuscript. The STEM videos also accompany the manuscript as supplemental materials.

**Table S1** Summary of parameters used during acquisition of STEM data.

| Video Title | Figure | Dwell Time [µs] | FOV [nm] | Dose/frame [# electrons] | Frame time [s] |
|---|---|---|---|---|---|
| MAADF_2b-d.avi | 2(b-d) | 32 | 128 | 7.87E+08 | 2.1 |
| MAADF_2e-g.avi | 2(e-g) | 63 | 128 | 1.54E+09 | 4.1 |
| MAADF_3b-d.avi | 3(b-d) | 63 | 128 | 1.54E+09 | 4.1 |
| MAADF_3e-g.avi | 3(e-g) | 63 | 128 | 1.54E+09 | 4.1 |
| MAADF_4b-d.avi | 4(b-d) | 32 | 64 | 7.87E+08 | 2.1 |

# Diffusion and E-beam Deposition Model

## *Simulation and Model Details*

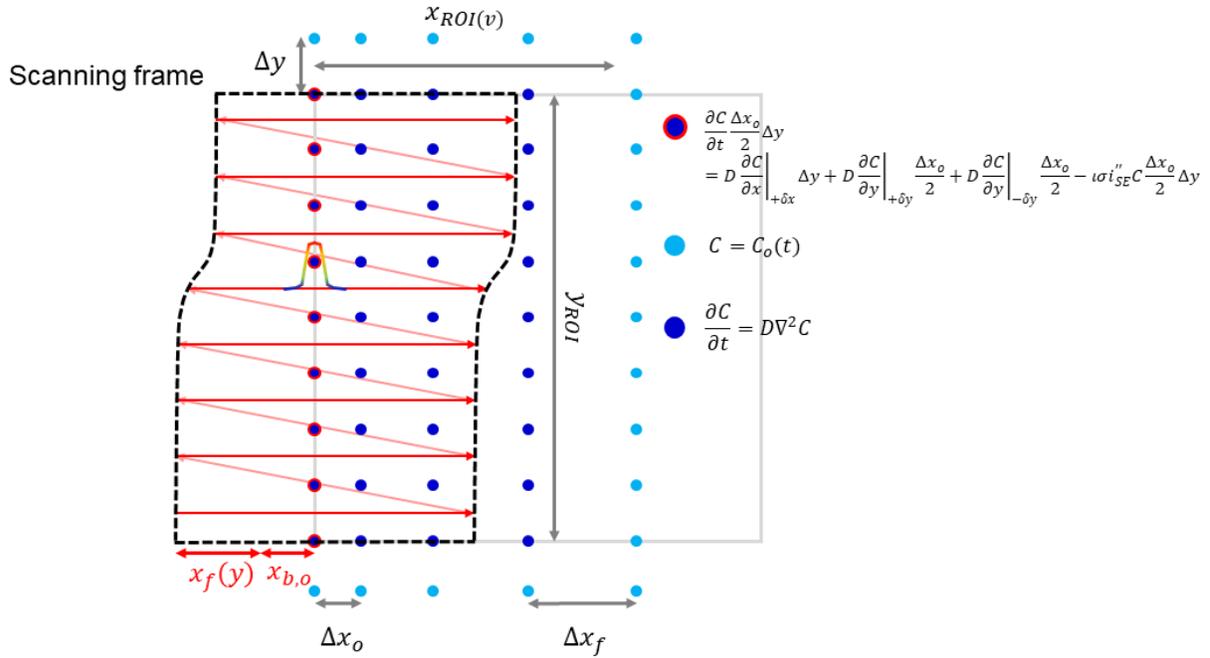

Figure S4: A schematic of the relationship between the computation pixel array (●), the real electron beam scanning frame (-) and the simulated beam scanning frame (--). The relationship between these features is described below.

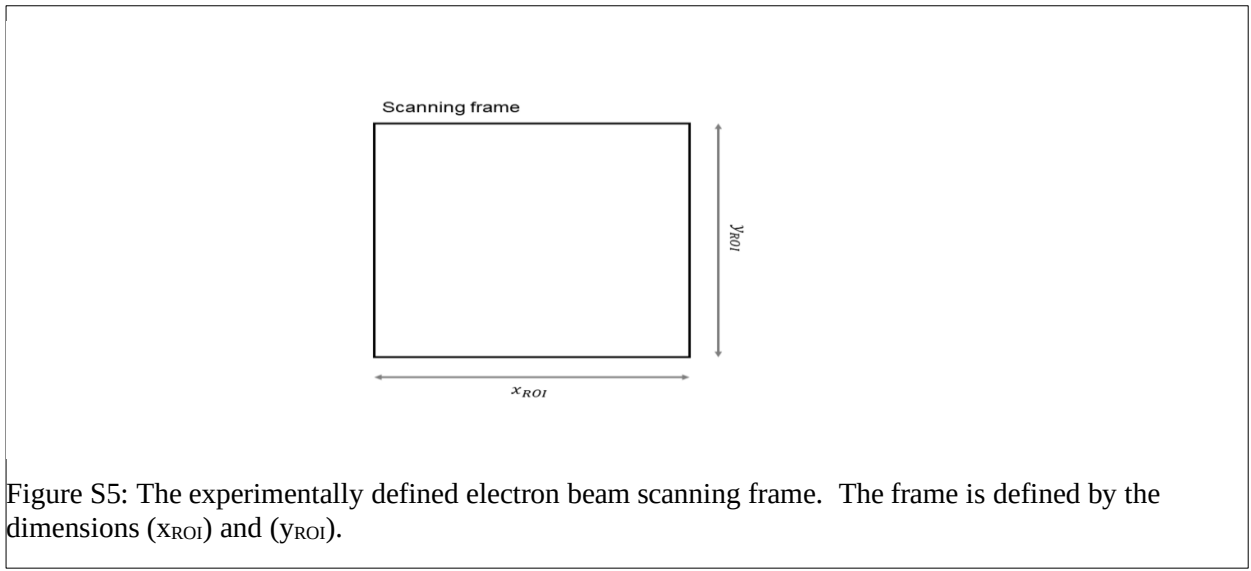

Figure S5: The experimentally defined electron beam scanning frame. The frame is defined by the dimensions ($x_{ROI}$) and ($y_{ROI}$).

## Fourier Number

The Fourier Number ($Fo_x$) is User defined input. The diffusion coefficient (D) and initial x–grid spacing is provided ($\Delta x_o$) are also User specified. This leaves one unknown in the time step ($\Delta t$).

$$\Delta t = \frac{Fo_x \Delta x_o^2}{D} [s] [1]$$

Fo$_x$ = 1 thus represents the characteristic time required for surface bound precursor to traverse a single pixel by diffusion. Fo$_x$ should be kept (< 1) to ensure that the contribution of precursor surface diffusion is not underestimated. The time step calculated from the User defined (Fo$_x$) is tested to make sure that the resolution is adequate to sample the various time–dependent processes that are relevant to the physical chemistry associated with the problem. This process is summarized below in the section 'Simulation time step conditioning'. The (Fo$_x$) is then updated to reflect the changes and reported to the User.

## Node spacing in the {x}–coordinate

A variable node spacing in the {x}–coordinate makes it possible to force a relatively small node spacing at the film-substrate interface, on the order of the primary electron beam size, while expanding to significantly larger values at large {x}–coordinate, where the surface precursor concentration gradients expected are small. The number of pixels in the {x}–coordinate is thus minimized reducing the total simulation time. The node spacing in the {x}–coordinate is determined by

$$\Delta x(x) = \Delta x_o + (\Delta x_f - \Delta x_o)(1 - e^{\frac{-x}{x_\tau}})[nm] \quad [2]$$

The {y}–coordinate spacing is constant in the simulation.

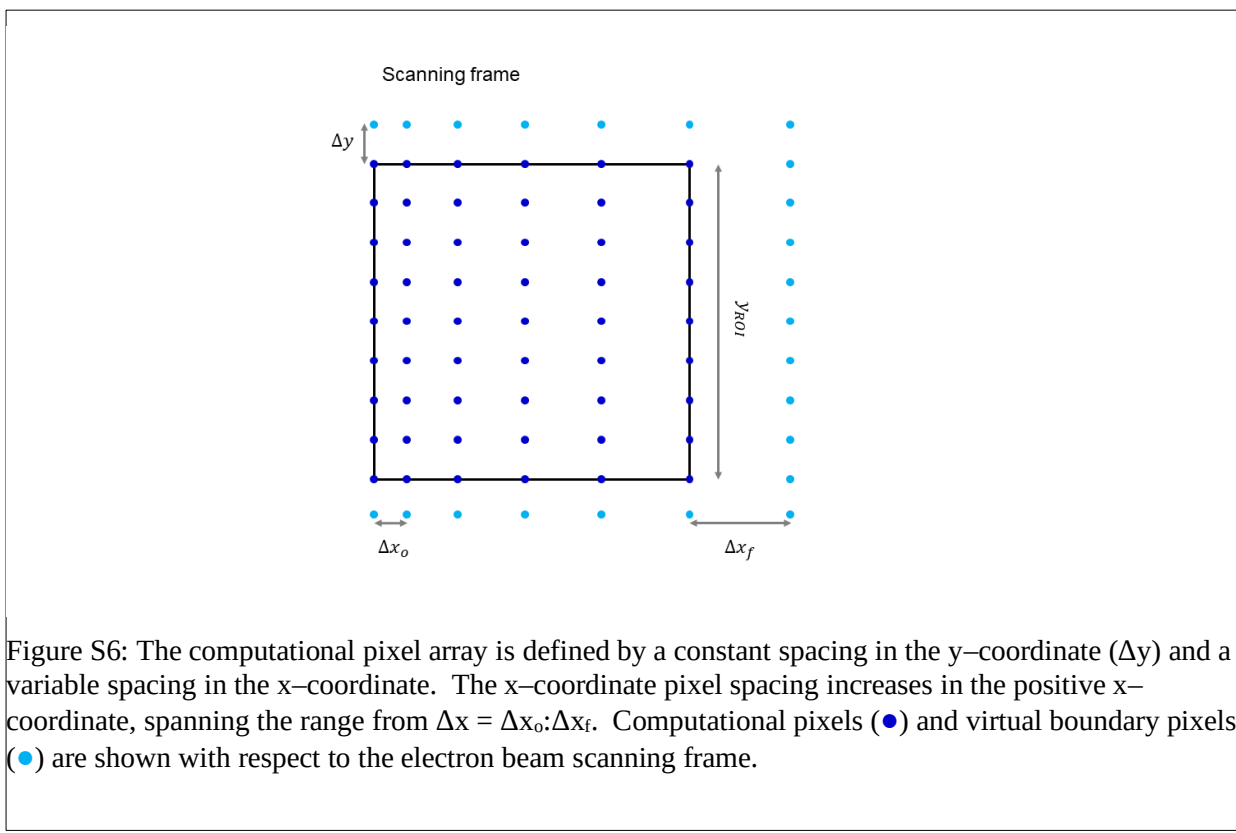

Figure S6: The computational pixel array is defined by a constant spacing in the y–coordinate ($\Delta y$) and a variable spacing in the x–coordinate. The x–coordinate pixel spacing increases in the positive x–coordinate, spanning the range from $\Delta x = \Delta x_o : \Delta x_f$. Computational pixels (●) and virtual boundary pixels (●) are shown with respect to the electron beam scanning frame.

*Simulation time step conditioning*

Multiple processes potentially limit the simulation time step to a maximum value. The smallest of these characteristic times determines the actual time step applied during the simulation. Integration of the PDE governing precursor surface transport during the pixel dwell time ($t_d$) must be sampled multiple times to accurately capture time–dependent changes in the precursor surface concentration inside the pixel – in principle ($Fo_x$) ensures proper time sampling. However, the steep concentration gradients that may be induced by the small STEM probe can generate relatively large gradients in both ($dC/dx$) and ($dC/dy$). The time sampling increment is set equal to

$$\Delta t \leq \frac{\tau_d}{8} [s] \quad [3]$$

where ($t_d$) is the dwell time per pixel. The pixel dwell time ($\tau_d$) is a User defined parameter. The simulation time step is next tested against a Scribe time ($\tau_{Sc}$). The scribe time is defined as the time required for the film/substrate interface to traverse the pixel width ($\Delta x_o$). As will be described below, the film/beam interaction will always occur at the left most x–coordinate pixel, with spacing ($\Delta x_o$). The scribe time is derived from the steady–state balance between the precursor dissociation rate and the film growth rate. The precursor dissociation rate ($dn_p/dt$) at the half pixel located at the film/substrate interface [molecules/s] is

$$\frac{dn_p}{dt} = \iota \sigma i''_{SE} C \frac{\Delta x_o}{2} \Delta y \left[\frac{molecules}{s}\right] \quad [4]$$

where ($i$) is the number of molecules dissociated per electron [molecules/e⁻], ($s$) is the electron impact dissociation cross–section [nm$^2$], ($i_{SE}$") is the electron beam flux [e⁻/nm$^2$ s], (C) is the precursor surface concentration [molecules/nm$^2$] and ($\Delta x_o \Delta y/2$) is the surface area of the pixel located in contact with the growing film [nm$^2$]. Please note that ($i_{SE}$'') appears in the dissociation term which is the secondary electron flux at the film edge which has a standard deviation of (a = 2 nm), as opposed to the actual STEM probe size of (a ~ 0.1nm), because the model assumes that the SEs generated just inside the edge of the film drive dissociation. The value of 2 nm is on the order of the mean free path of an SE in carbon.[7] Also, since the secondary electron yield for the carbon–based film is unknown, the SE current density is taken simply as proportional to the beam current while the SE range is explicitly taken account of through (a).

As stated previously, the dissociation rate is in equilibrium with the film growth rate ($dn_f/dt$) and the film deposition rate can be expressed in term of the lateral film velocity, or

$$\frac{dn_p}{dt} = \frac{dn_f}{dt} = C_f \frac{dx}{dt} \Delta y z_o \left[\frac{molecules}{s}\right] \quad [5]$$

where ($C_f$) is the film concentration, i.e., density, ($dx/dt$) is the lateral film growth velocity and ($z_o$) is the film thickness which is taken as the molecular size ~0.3 nm. The scribe time is derived by combining equations 4 & 5, substituting $t_{Sc}$ for ($dt$) and $\Delta x_o$ for ($dx$), and solving for $t_{Sc}$

$$\tau_{Sc} = \frac{2C_f z_o}{\sigma \frac{i_b}{2\pi a^2} \theta s_d} [s][6]$$

The simulation time step (Δt) must be less than the scribe time. The following condition is used in the simulation

$$\Delta t \leq \frac{\tau_{Sc}}{8} [s][7]$$

Otherwise, the film/substrate interface will advance at rate that may be changing significantly during the time step (Δt) leading to a poor sampling of the electron beam edge intensity. A final conditioning step is applied to make sure that the time step is an integer number of the pixel dwell time, or

$$\Delta t = \frac{\tau_d}{round|\frac{\tau_d}{\Delta t}|} [s][8]$$

## *Scanning frame {x} length (virtual)*

The equilibrium precursor surface concentration, i.e., the surface concentration under the beam off condition, is described by

$$C_o = \theta s_d [\frac{molecules}{nm^2}][9]$$

where (θ) represents the fraction (0–1) of the maximum possible precursor surface concentration as dictated by the binding site density ($s_d$). Often, exposure of the film interface with the electron beam leads to a depletion of (C) such that

$$C \ll C_o [10]$$

A relatively large diffusion coefficient (D) can lead to the nearly full recovery of C(x,y), or

$$C(x,y) \cong C_o [11]$$

before the electron beam reaches the line scan boundary position of x = $x_{ROI}$. Simulation time is thus wasted if the electron beam continues to the x–frame boundary for this 'full precursor refresh' condition. The program is set to detect this situation by comparison of the minimum concentration inside the scanning frame against a User specified parameter that represents full precursor refresh, e.g.,

$$\theta_r = 0.995\theta [0-1][12a]$$

by way of

$$\min_{x,y} \sum_{x=0}^{x_{ROI}} \sum_{y=0}^{y_{ROI}} C(x,y) > 0.995\theta s_d [\frac{molecules}{nm^2}][12b]$$

If the precursor refresh condition is met, the electron beam advances to the next line scan origin to begin growth. Also, as part of this special precursor refresh condition, the simulation time is advanced forward by the remaining line scan time to preserve the experiment and simulation time comparison. This technique is possible because film deposition only occurs as the beam passes over the film/substrate interface such that the total length of the scan is irrelevant, at least when the precursor refresh time is less than the total line scan time.

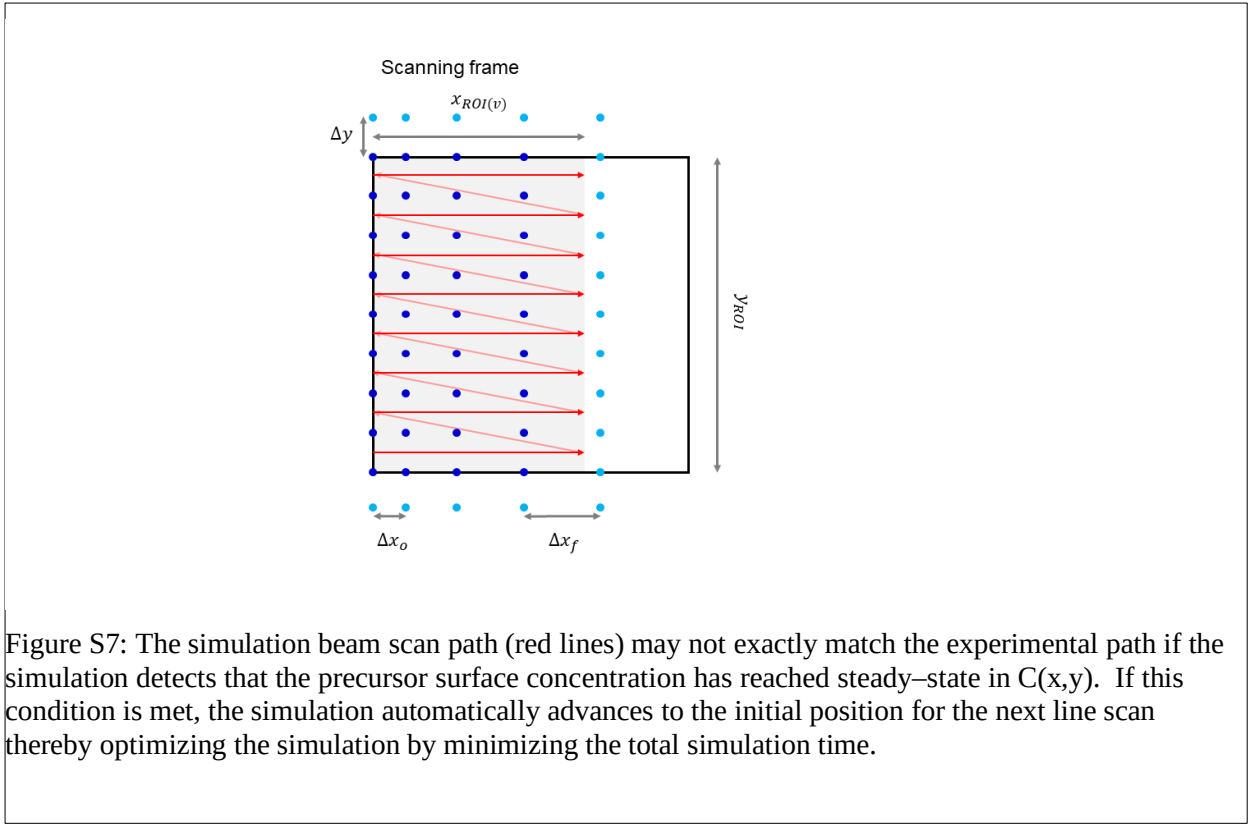

Figure S7: The simulation beam scan path (red lines) may not exactly match the experimental path if the simulation detects that the precursor surface concentration has reached steady–state in C(x,y). If this condition is met, the simulation automatically advances to the initial position for the next line scan thereby optimizing the simulation by minimizing the total simulation time.

The primary electron beam moves at the velocity defined by the User defined pixel point pitch ($\Lambda$) and the primary electron beam dwell time per pixel ($\tau_d$)

$$v_{b,x} = \frac{\Lambda}{\tau_d} [\frac{nm}{s}] [13]$$

The *real*, primary electron linear beam scanning time

$$t_{1D} = \frac{x_{ROI}}{\Lambda} \tau_d [s][14]$$

is used to calculate the real growth rate, even when the simulation implements the full precursor refresh condition described in the previous section. The real frame scanning time is

$$t_{2D} = (\frac{y_{ROI}}{\Lambda} + 1) t_{1D} [s][15]$$

and the total simulation time

$$t = P \cdot t_{2D} [s][16]$$

where (P) is the number of frame loops.

*Primary electron beam scanning*

Film/interface growth is modeled by shifting the origin of the primary electron beam in the negative x–coordinate per simulation time step (Δt). This model is, at least in part, validated by the fact that the interface advances with a near constant x–coordinate, independent of the y–coordinate. This fact is derived from experimental results. Conversely, the case of a propagating curved interface would invalidate such a model because the PDE describing C(x,y) would be 'blind' to the curvature. In the framework of this linear interface propagation model, the electron beam flux equation

$$i''_{SE}(x,y,t) = \frac{i_b}{2\pi a^2} e^{\frac{-(\Delta x_o - x_b(t))^2 - (m\Delta y - y_b(t))^2}{2a^2}}$$

$$n = 1$$

$$m = 1,2,3,\dots M$$

is calculated only along the first column (m = 1) of the discretized array of computational pixels which remains effectively fixed to the interface throughout the simulation. Thus, film deposition, and the concomitant interface displacement in the positive {x} direction, is captured by displacing the beam in the negative {x} direction relative to the spatial node array. The electron beam displacement equations

$$x_b(t) = x_{b,o} - x_f(y,t) + v_{b,x} t [nm] \quad [18]$$

$$y_b(t) = x_{b,o} + v_{b,y} t [nm] \quad [19]$$

require several terms to implement the displacement.

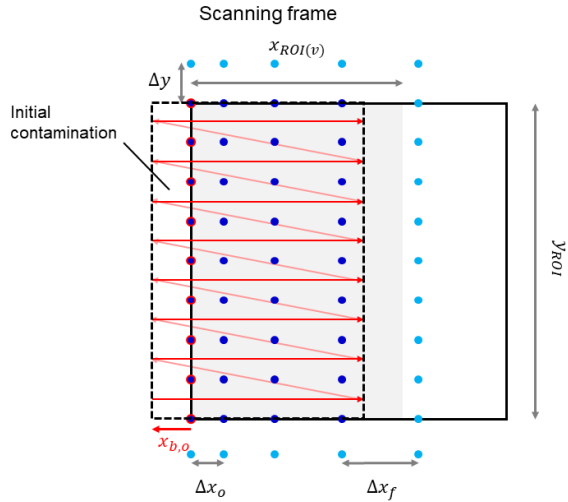

Figure S8: Initially the beam origin is displaced in the negative x–coordinate by ($x_{b,o}$) to emulate real experiments where the beam intersects the surrounding boundary contamination to initiate deposition during the first frame scan. A single line scan is completed at position $x_{ROI} - x_{b,o} - x_{f(y,t)}$ to conserve the frame scan time. $x_{f(y,t)}$ accounts for any growth during the line scan and is updated every simulation time step ($\Delta t$).

The variable ($x_{b,o}$) represents the starting position of the beam, at t = 0, relative to the preexisting contamination on the surface. By default, the initial x beam position is shifted sufficiently in the negative direction to ensure at least a complete interaction of the primary electron beam profile with the initial contamination interface occurs during the initial irradiation frame.

$$x_{b,o} = -2 \cdot FWHM[nm] [20]$$

and FWHM is related to the standard deviation of the beam profile, or (a), by way of a ~ FWHM/2.355. The variable $x_f(y,t)$, updated per Dt, tracks the interface growth at the m = 1 column of pixels.

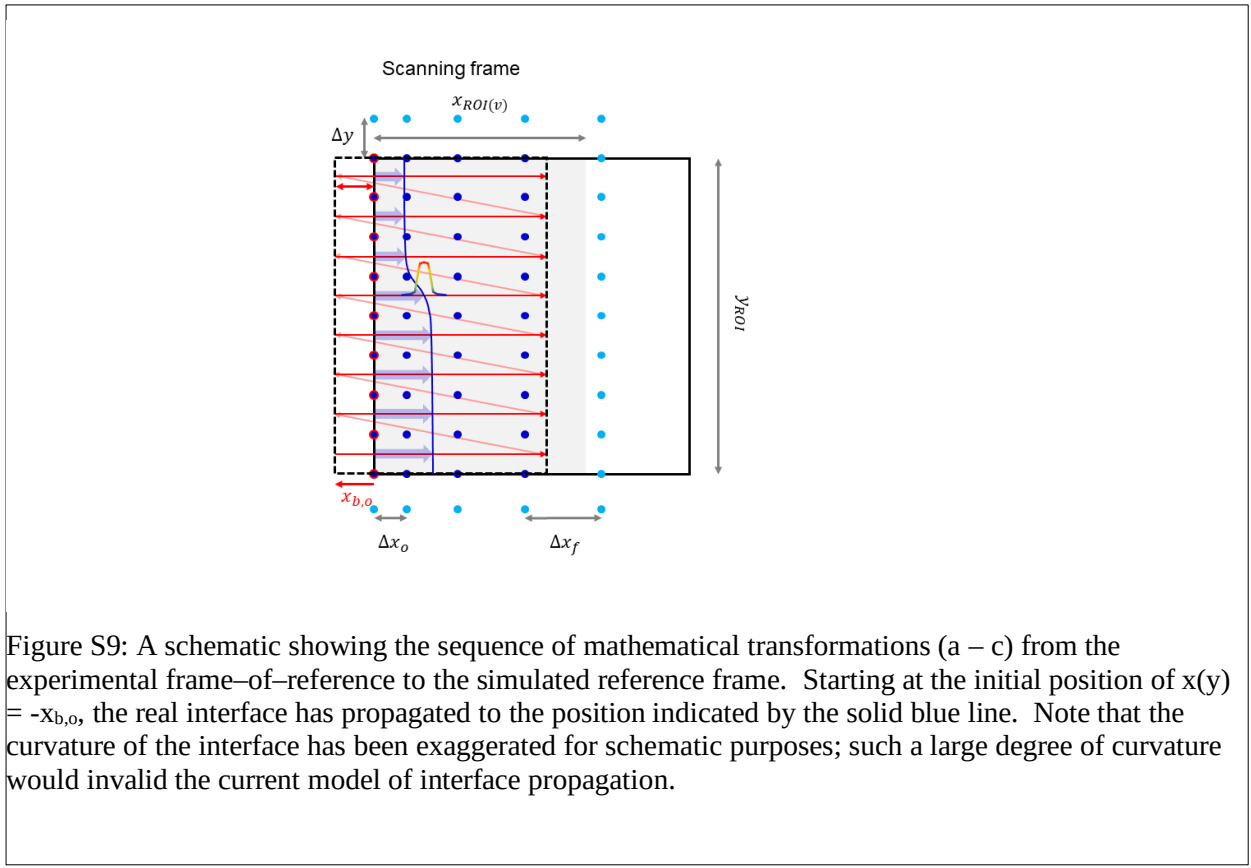

Figure S9: A schematic showing the sequence of mathematical transformations (a – c) from the experimental frame–of–reference to the simulated reference frame. Starting at the initial position of x(y) = -$x_{b,o}$, the real interface has propagated to the position indicated by the solid blue line. Note that the curvature of the interface has been exaggerated for schematic purposes; such a large degree of curvature would invalid the current model of interface propagation.

Consider the situation within the ROI, as shown in figure S9, where the interface has advanced to the position shown by the solid blue line following many frames of deposition. The current (x,y) position of the Gaussian beam is also shown to indicate the degree of beam/interface overlap required to induce deposition. In a complementary approach, the film/interface position can be implicitly treated by imposing a transient negative {x} beam shift, per line $x_f(y)$ (figure S10), to affix the interface position always at x = 0, or

$$x_f(y, t) = 0 \quad [21]$$

The hatched blue line (figure S10) shows the origin in the current example for each line in the frame due to the transformation.

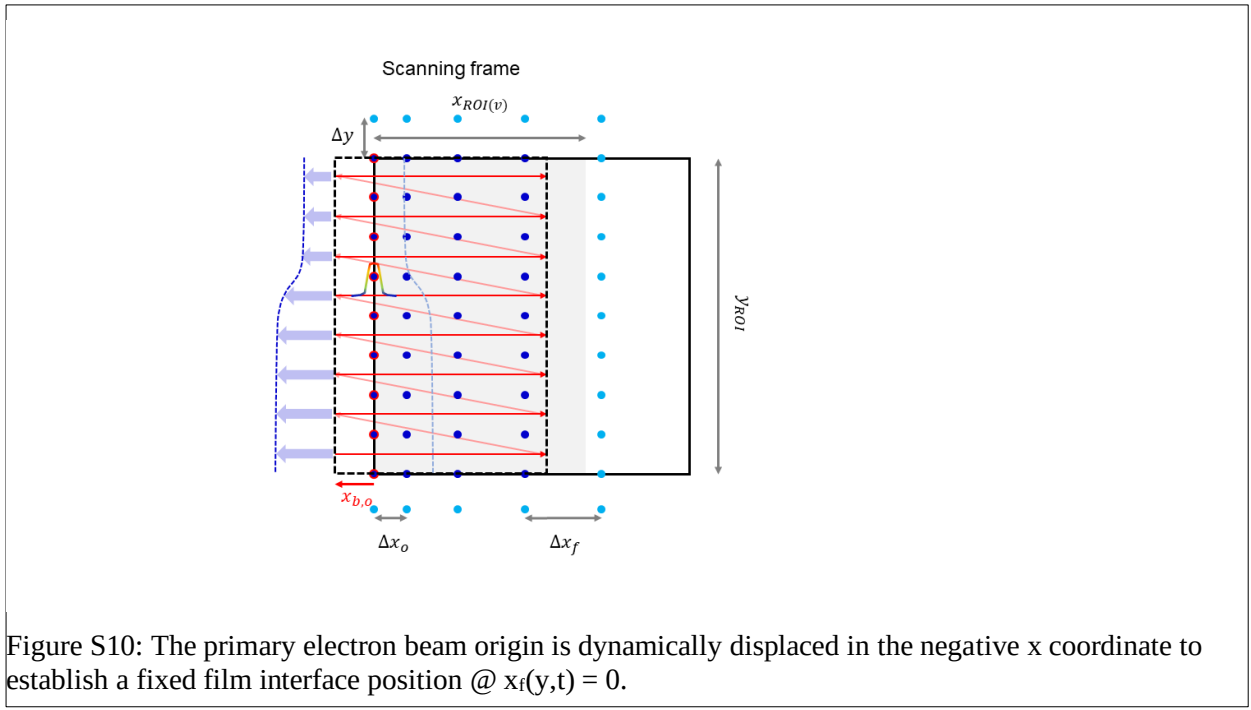

Figure S10: The primary electron beam origin is dynamically displaced in the negative x coordinate to establish a fixed film interface position @ $x_f(y,t) = 0$.

Ultimately, the outcome of the spatial reference frame transformation is (1) a fixed computational coordinate system with (2) a dynamically distorted ROI (black hatched line, figure S11) that exactly replicates the square frame experimental scan, at least in the limit of a nearly linear film interface

$$\frac{dx_f}{dy} \approx 0 \qquad [22]$$

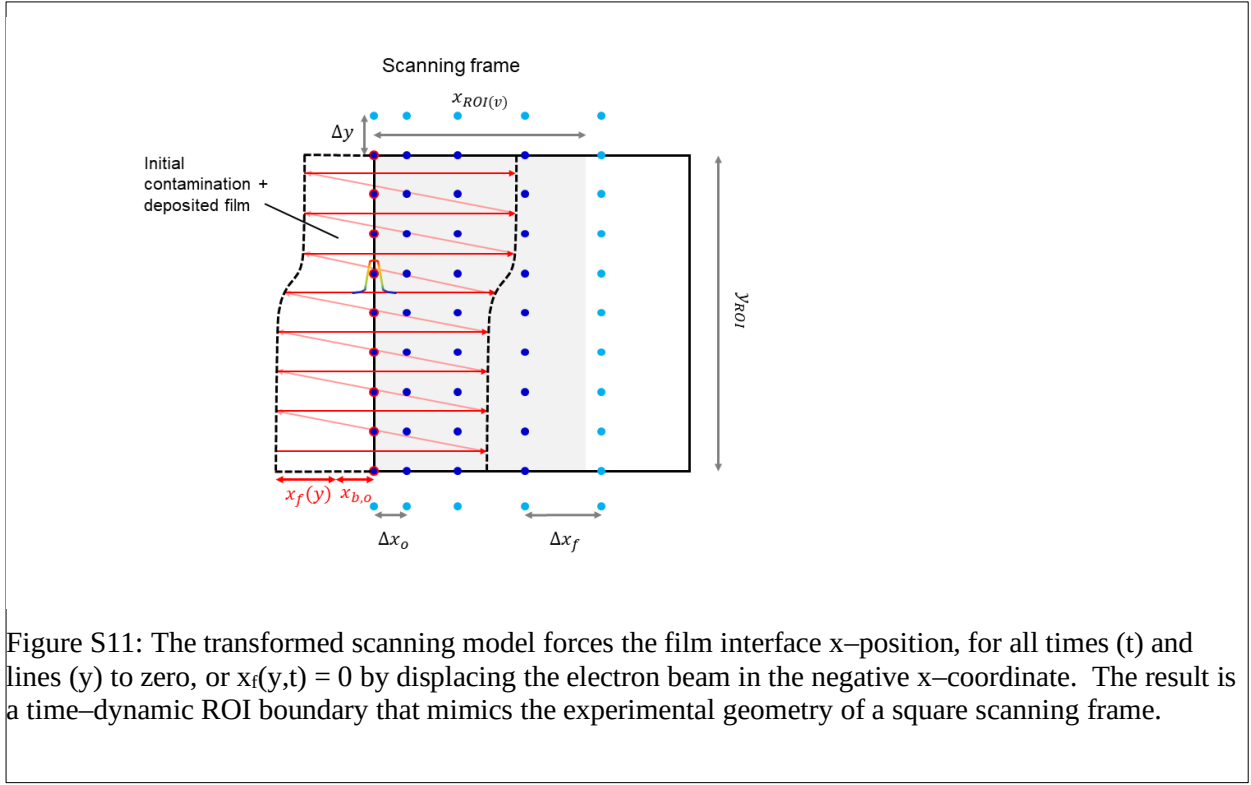

Figure S11: The transformed scanning model forces the film interface x–position, for all times (t) and lines (y) to zero, or $x_f(y,t) = 0$ by displacing the electron beam in the negative x–coordinate. The result is a time–dynamic ROI boundary that mimics the experimental geometry of a square scanning frame.

## *Gradient Descent*

A gradient descent was executed, in the two variables (D) and (θ), to attempt to identify a unique solution for each STEM experiment. The gradient descent is performed on an error function relative to real experiments

$$f(D, \theta) = (v_f - v_{f,\exp})^2 [\frac{nm^2}{s^2}][23]$$

where ($v_f$) is the lateral film growth velocity and ($v_{f,\exp}$) is the experimental value. An initial guess is made for each independent variable

$$D = D_o[\frac{nm^2}{s}][24a]$$

$$\theta = \theta_o[0-1][24b]$$

Derivatives will be required to determine the gradient in each variable. The gradient is estimated using a finite difference approach and the range over which the gradient is calculated is based on the initial guess

$$\Delta D = 0.2 D_o [\frac{nm^2}{s}][25a]$$

$$\Delta \theta = 0.2 \theta_o [0-1][25a]$$

The following description provides a sequential summary of a one loop iteration of the gradient descent

**for**

The algorithm begins with simulation execution at the node

$$f(D_o, \theta_o)[step1]$$

The gradient descent is executed, first in the variable (D), and next in the variable (q). The gradient estimate in (D) requires the simulation at two additional parameter space nodes

$$f(D_o + \Delta D, \theta_o)[step2]$$

$$f(D_o - \Delta D, \theta_o)[step3]$$

making it possible to solve for df/dD

$$\frac{df}{dD} \cong \frac{f(D_o + \Delta D, \theta_o) + f(D_o - \Delta D, \theta_o) - 2f(D_o, \theta_o)}{2\Delta D}[step4]$$

The learning rule is based on the initial gradient calculation and the initial guess in the independent parameter ($D_o$)

$$\alpha_D = \frac{D_o}{\frac{df}{dD}|_o}[\frac{nm^2}{s}][step5]$$

and is used to update the value of (D)

$$D_{m+1} = D_m - \alpha_D \frac{\partial f}{\partial D}[\frac{nm^2}{s}][step6]$$

and (m) is the iteration variable. Simulation execution gives a better estimate of the function minimum at

$$f(D_{m+1}, \theta_o)[step7]$$

The gradient estimate in (q) requires the simulation at two additional parameter space nodes

$$f(D_{m+1}, \theta_o + \Delta\theta)[step8]$$

$$f(D_{m+1}, \theta_o - \Delta\theta)[step9]$$

making it possible to solve for df/dθ

$$\frac{df}{d\theta} \cong \frac{f(D_{m+1}, \theta_o + \Delta\theta) + f(D_{m+1}, \theta_o - \Delta\theta) - 2f(D_{m+1}, \theta_o)}{2\Delta\theta}[step10]$$

The learning rule is based on the initial gradient calculation and the initial guess in the independent parameter

$$\alpha_\theta = \frac{\theta_o}{\frac{df}{d\theta}|_o}[0-1][step11]$$

and is used to update the value of (D).

$$\theta_{m+1} = \theta_m - \alpha_\theta \frac{\partial f}{\partial \theta} [0-1][step12]$$

The function after a single loop is finally

$$f(D_{m+1}, \theta_{m+1})[step13]$$

**end**

## Supporting Information 2

*Table of Parameters and Descriptions*

| | |
|---|---|
| $D = 10^3 - 10^6 [nm^2/s]$ | Surface diffusion coefficient |
| $\sigma = 0.02 [nm^2]$ | Electron impact dissociation cross–section |
| $i_b = 60 [pA]$ | Electron probe current |
| $FWHM = 2 [nm]$ | Electron probe size of 0.125 nm changed to 2 nm based on SE MFP) |
| $s_d = 60 [sites/nm^2]$ | Graphene binding site density (estimated from STEM image of graphene) |
| $\theta = 10^{-6} - 10^{-3} [0 - 1]$ | Maximum fractional precursor surface coverage |
| $C_o = \theta s_d [sites/nm^2]$ | Initial surface concentration |
| $C_f = 25 [molecules/nm^3]$ | Deposit density (0.5 g/cm³ based on 12 g/mol) |
| $z_o = 0.3 [nm]$ | Deposit film thickness (characteristic monolayer thickness) |
| $\Delta x_o, x_\tau, \Delta x_f = 0.125, 36, 12 [nm]$ | Simulation min step, decay constant and max step |
| $\Delta y = 0.125 [nm]$ | Simulation pixel size in the y–dimension |
| $x_{b,o} = -2 \cdot FWHM [nm]$ | Beam x–displacement relative to initial interface position |

*Experiments 3, 4, 5 and 6*

| | |
|---|---|
| $PoP = 0.5, 0.25, 0.0625, 0.125 [nm]$ | Pixel point pitch |
| $\tau = 64, 32, 16, 32 [\mu s]$ | Pixel dwell time |
| $FOV = 128, 64, 32, 32 [nm]$ | Field–of–view |
| $v_f = 0.045, 0.045, 0.0076, 0.015 [nm/s]$ | Film growth velocity (experimental) |

*Parameters (known & unknown)*

*Known Variables*

*Deposited film thickness*

$$z_o = 0.3 nm$$

The characteristic molecular size was assumed as the deposited film thickness in the z–dimension.

*Carbon film concentration*

$$C_f = \frac{0.5[\frac{g}{cm^3}] 6.02 x 10^{23}[\frac{molecules}{mol}] 1 x 10^{-21}[\frac{cm^3}{nm^3}]}{12[\frac{g}{mol}]} \cong 25[\frac{molecules}{nm^3}]$$

Density on the order of a porous amorphous/glassy carbon.

*Surface binding site density*

$$s_d \cong 60[\frac{atoms}{nm^2}]$$

Estimated from the clean graphene image presented as figure 1(b) in the manuscript.

## *Unknown Variables*

*Initial & boundary precursor surface concentration*

$$C_o = \theta s_d$$

We assume a relatively low concentration of precursor commiserate with predicted deposition rates observed experimentally

$$\theta = 1 x 10^{-6} - 1 x 10^{-3}$$

*Electron impact dissociation cross-section*

$$\sigma = 2 x 10^{-2}[nm^2]$$

The relatively large estimate reflects the higher probability of dissociation associated with a relatively large hydrocarbon species that can remain on the surface at elevated temperatures.

Precursor surface diffusion coefficient

$$D = 1 x 10^3 - 1 x 10^6 [\frac{nm^2}{s}]$$

This range of value of (D) gave experimental results on the order of experiments as shown in the results below.

## Supporting Information 3

*Evidence of Total Precursor Pressure Decrease*

The simulation space maps below show that the path of solutions is consistent across all four experiments, if, a shift of (-Δθ) is applied to the solution curve. The magnitude of the shift is proportional to the total time the sample has been evacuated in the STEM chamber.

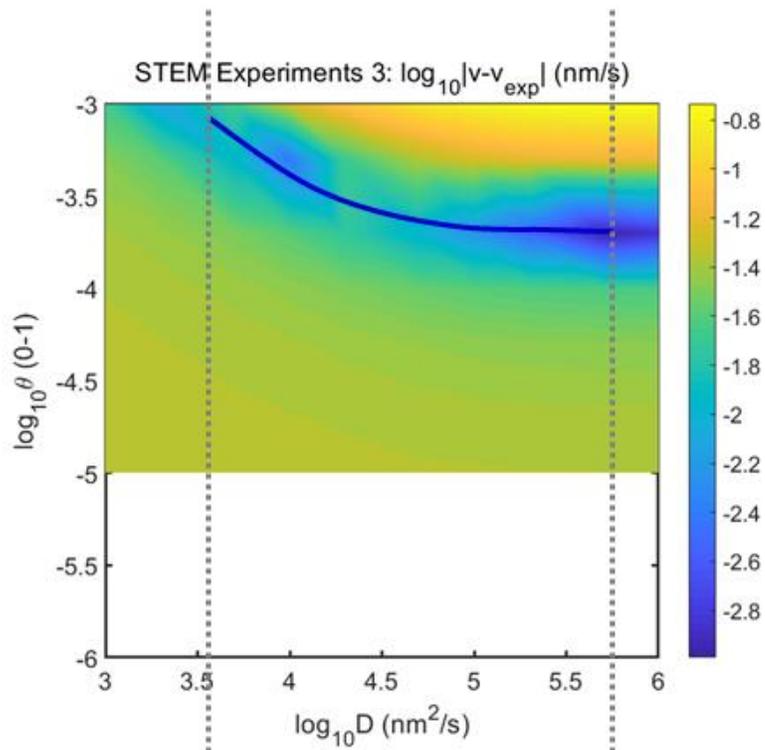
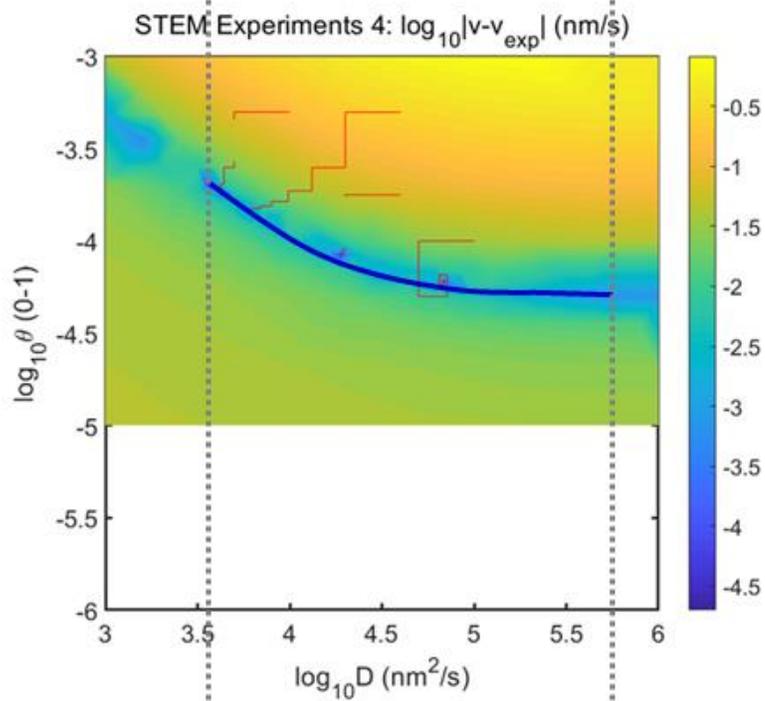

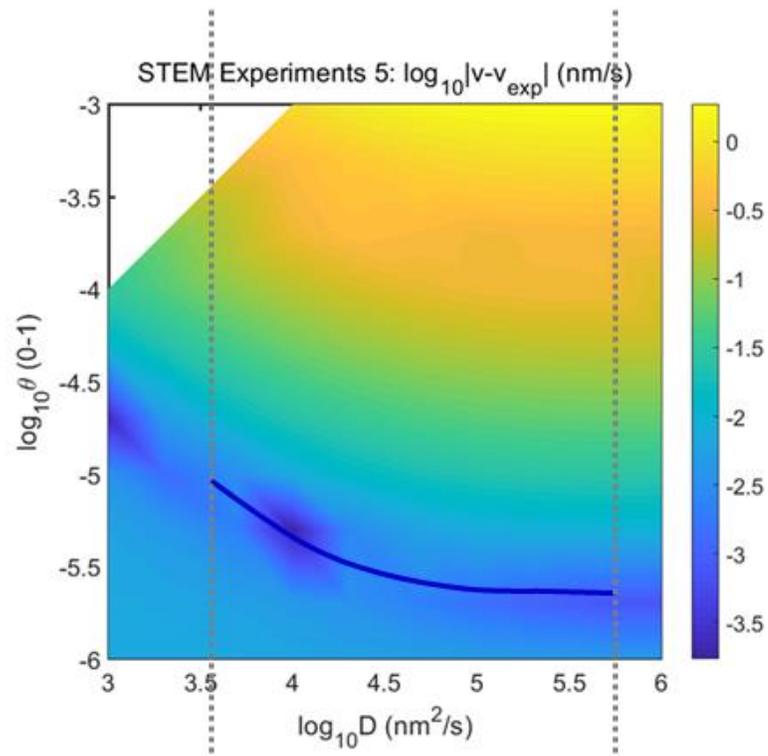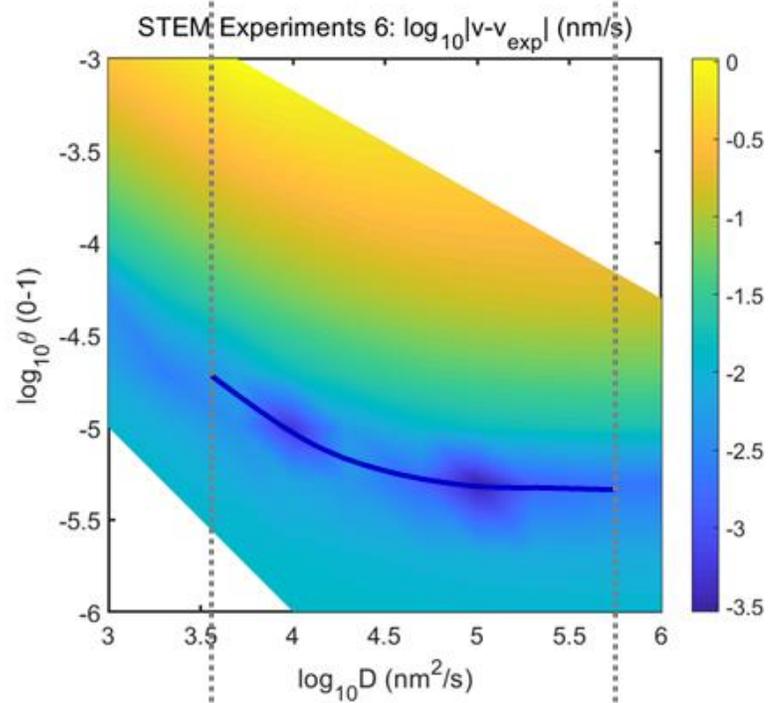